\documentclass[12pt]{article}
\pdfoutput=1

\usepackage{amssymb,bm}

\usepackage{xcolor}
\usepackage[colorlinks]{hyperref}
\hypersetup{
  colorlinks=true,
  linkcolor=purple,
  filecolor=red,
  citecolor=blue,      
  urlcolor=violet
}

\usepackage{scicite}
\usepackage{graphicx}
\usepackage{times}
\usepackage[normalem]{ulem}
\usepackage{soul}
\usepackage{multirow}
\usepackage{array}
\usepackage{booktabs}
\usepackage{lineno}

\usepackage{subfig}
\usepackage[font=footnotesize]{caption}

\graphicspath{images/}

\usepackage[symbol]{footmisc}

\graphicspath{{\subfix{./images/}}}

\newenvironment{sciabstract}{%
\begin{quote} \bf}
{\end{quote}}

\title{A tera-electronvolt afterglow  from a narrow jet in an extremely bright gamma-ray burst 221009A }

\author{}
\date{}

\begin{document}

\baselineskip 20pt


\maketitle

\centerline{\author{\large LHAASO Collaboration\footnotemark[1]\footnotemark[2]}}
\vspace{10pt}

\footnotetext[1]{Corresponding authors:  X.Y.~Wang (xywang@nju.edu.cn), Z.G.~Yao (yaozg@ihep.ac.cn), \\
\indent\indent Z.G.~Dai (daizg@ustc.edu.cn), M.~Zha (zham@ihep.ac.cn), Y.~Huang (huangyong96@ihep.ac.cn), \\
\indent\indent J.H.~Zheng (mg21260020@smail.nju.edu.cn)}
\footnotetext[2]{The LHAASO Collaboration
authors and affiliations are listed in the supplementary materials}

\begin{sciabstract}
Some gamma-ray bursts (GRBs) have an afterglow in the tera-electronvolt (TeV) band, but the early onset of this afterglow has not been observed. We report observations with the Large High Altitude Air Shower Observatory of the bright GRB 221009A, which serendipitously occurred within the instrument field of view.
More than 64,000 photons (above 0.2~TeV) were detected within the first 3000 seconds. The TeV photon flux began several minutes after the GRB trigger, then rose to a flux peak about 10 seconds later. This was followed by a decay phase, which became more rapid  at $\bm{\sim 650\,{\rm s}}$ after the peak. The emission can be explained with a relativistic jet model with half-opening angle  $\bm{\sim 0.8^\circ}$, consistent with the core of a structured jet.  This interpretation could  explain  the high isotropic energy of this GRB.

\end{sciabstract}


\section*{}
Gamma-ray bursts are violent explosions observed in distant galaxies, characterized by a rapid flash of gamma rays lasting from a fraction of a second up to several hundreds of seconds. The progenitors of the long-duration ($\gtrsim 2$ seconds)  GRBs are thought to be collapsing massive stars, while those of short-duration ($\lesssim 2$ seconds) ones are  the merger of two compact objects~\cite{GRB-review}. The emission of a GRB consists of two stages, the prompt emission and the afterglow, and the two stages can partially overlap in time. The prompt emission has irregular variability on timescales as short as milliseconds, which is thought to result from internal shocks or other  dissipation mechanisms that occur at small radii. The afterglow emission is much more smooth and lasts much longer. The flux usually shows power-law decays in time. The long-lasting afterglow  is thought to  result from external shocks caused by the interaction of the relativistic jets with the ambient medium at large radii. The afterglow emission spans a wide range of frequencies of electromagnetic wave. The radio  to
sub-GeV emission  is generally interpreted as synchrotron radiation from relativistic electrons that are
accelerated by the external shock~\cite{GRB-review}. It has been predicted that the same electrons up-scatter the synchrotron photons via the inverse Compton (IC) mechanism, producing a synchrotron self-Compton (SSC) emission component extending into the very high energy (VHE, $>100$~GeV) regime~\cite{Sari&Esin2001,Wang2001,Zhang2001,Zou2009}. 

VHE emission has been detected from a few GRBs  during the afterglow decay phase~\cite{190114C-1,190114C-2,180720B,190829A}. However, those observations used  pointed instruments that needed to slew to the GRB position, so did not capture the prompt and  afterglow rising phases. 
Extensive air shower detectors have  larger instantaneous field of view,  near $100\%$ duty cycle and do not need to be pointed, so could potentially observe
the prompt GRB and afterglow onset phases. However, attempts to observe
GRBs with  extensive air shower detectors  have not resulted in detections~\cite{HAWC1,ARGO-prompt,HAWC2}.

\section*{Observations  of GRB 221009A}

On 9 October 2022, at 13:16:59.99 Universal Time (UT) (hereafter $T_0$), the  Gamma-Ray
Burst Monitor (GBM) on the Fermi spacecraft detected and located the burst  GRB 221009A~\cite{GCN-GBM}. The Large Area Telescope (LAT) on Fermi also detected high-energy emission from the burst~\cite{GCN-LAT}. The  Burst Alert Telescope (BAT) on the Neil Gehrels Swift Observatory (Swift) spacecraft detected this burst 53 minutes later, and Swift's X-Ray Telescope (XRT) observations began 143 seconds after
the BAT trigger~\cite{GCN-BAT}.  The exceptionally large fluence of this event saturated almost all gamma-ray detectors during the main burst.  The event fluence of GRB 221009A was measured by the Konus-Wind spacecraft from $T_0+175$~s to $T_0+1458$~s
as $\gtrsim  5\times 10^{-2}\,{\rm  erg\,cm^{-2}}$ in the energy range 10 to 1000~keV~\cite{GCN-KW}, higher than any previously observed GRB. Later optical observations measured the redshift ($z$) of the afterglow, finding  $z = 0.151$~\cite{GCN-redshift,GCN-redshift2}. Assuming standard cosmology,  the burst isotropic-equivalent energy release $E_{\rm \gamma,iso}$ is  at least $3\times 10^{54}$~erg, among the 
highest  measured.

We observed GRB 221009A with the Large
High Altitude Air Shower Observatory (LHAASO). At the time of the GBM trigger, GRB~221009A was within the field of view of LHAASO, at a zenith angle of $28.1^{\circ}$ (Fig.~\ref{fig:wcda-fov}), and was observed by LHAASO  for about 6000~s  before moving out of the field of view. LHAASO is a VHE gamma-ray extensive air shower detector~\cite{LHAASO}, consisting of three interconnected detectors. We analyze the observations of GRB 221009A with LHAASO's Water Cherenkov Detector Array
(WCDA, Fig.~\ref{fig:WCDA-fig})~\cite{LHAASO}, which detected the burst at coordinates of right ascension (RA)
 $288.295 \pm 0.005$ (stat) $\pm 0.05$ (sys) degrees and declination  $19.772 \pm 0.005$ (stat) $\pm 0.05$ (sys) degrees, with a statistical significance 
$>250$ standard deviation (Fig.~\ref{fig:sigmap}).  Within the first $\sim 3000\,{\rm s}$ after the trigger, WCDA detected more than 64,000 photons with energies between $\sim 200\,{\rm GeV}$ and  $\sim 7\,{\rm TeV}$ associated with the GRB. At these energies, another detector reported no emission about 8 hours after $T_0$~\cite{HAWC-221009A}.


\begin{figure}[b]
\centering
\includegraphics[width=0.95\linewidth]{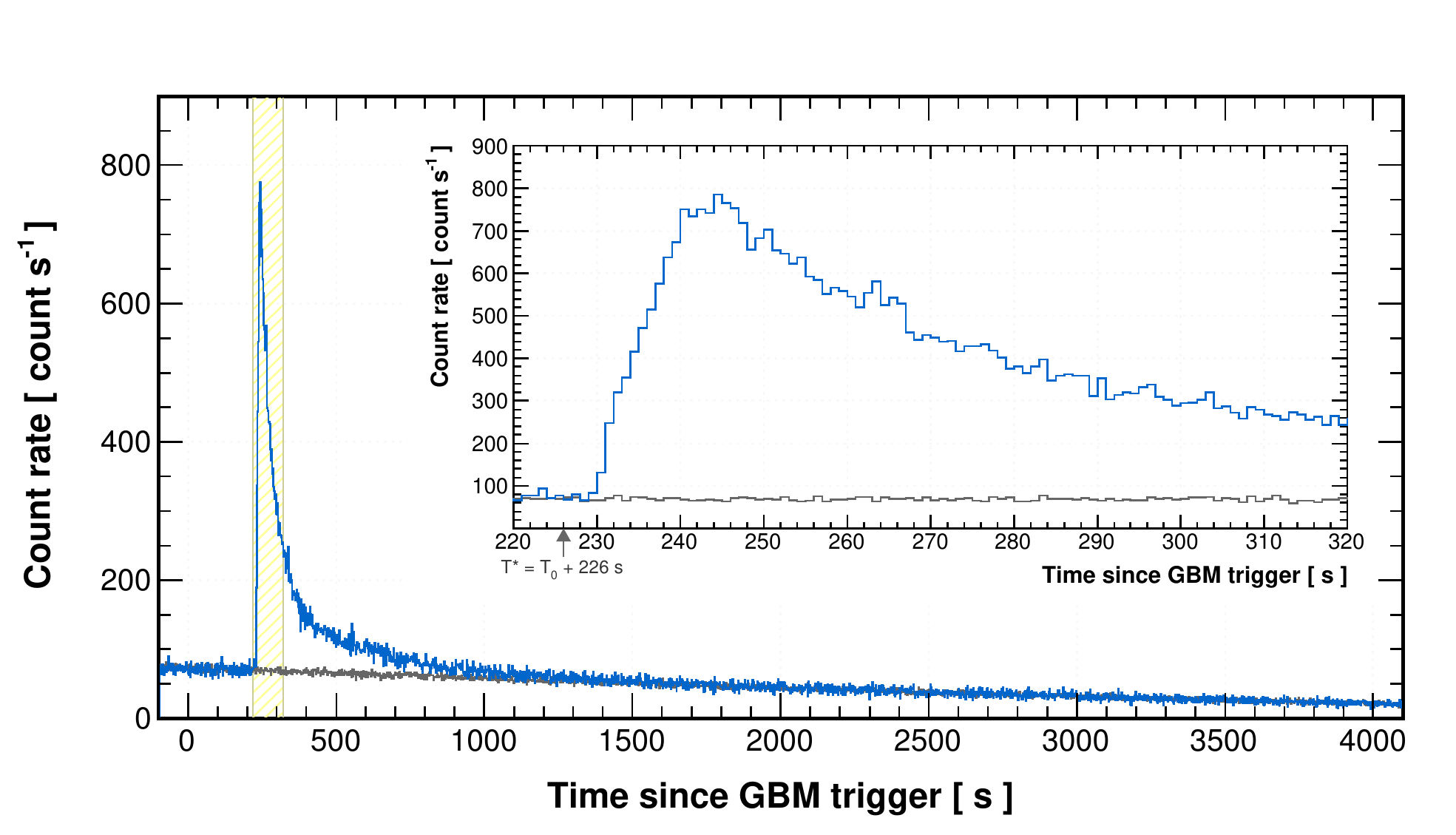}
\caption{{\bf Count-rate Light Curve of GRB 221009A Observed by LHAASO-WCDA.} The energy range of photons observed is around 0.2--7~TeV. The inset panel shows a zoomed-in view of the light curve during 220--320~s (shaded zone in light yellow) after the GBM trigger ($T_0$), with the arrow indicating the reference time $T^{*}=T_0+226\,{\rm s}$ for our light curve analysis (see text). Blue histograms are the data and black histograms are the estimated background.}
\label{fig:LC-linear}
\end{figure}

Fig.~\ref{fig:LC-linear} displays the WCDA light curve  (count rate as a function of time) of GRB 221009A. The VHE emission exhibits a smooth temporal profile, with a rapid rise to a peak, followed by a more gradual decay that persists for at least 3000~s after the peak. In contrast, the MeV gamma-ray light curves, measured by GBM~\cite{GCN-GBM} and other gamma-ray detectors~\cite{GCN-HEBS,GCN-KW}, are highly variable. The GBM emission includes an initial precursor pulse lasting approximately 10~s, which sets $T_0$, followed by an extended, much brighter, multi-pulsed emission episode (Fig.~\ref{fig:rate_gbm_wcda}). The contrast between the TeV and MeV light curves suggests that the TeV emission has a different origin than the prompt MeV emission.

\subsection*{Analysis of the gamma-ray data}

\begin{figure}
\centering
\includegraphics[width=0.9\linewidth]{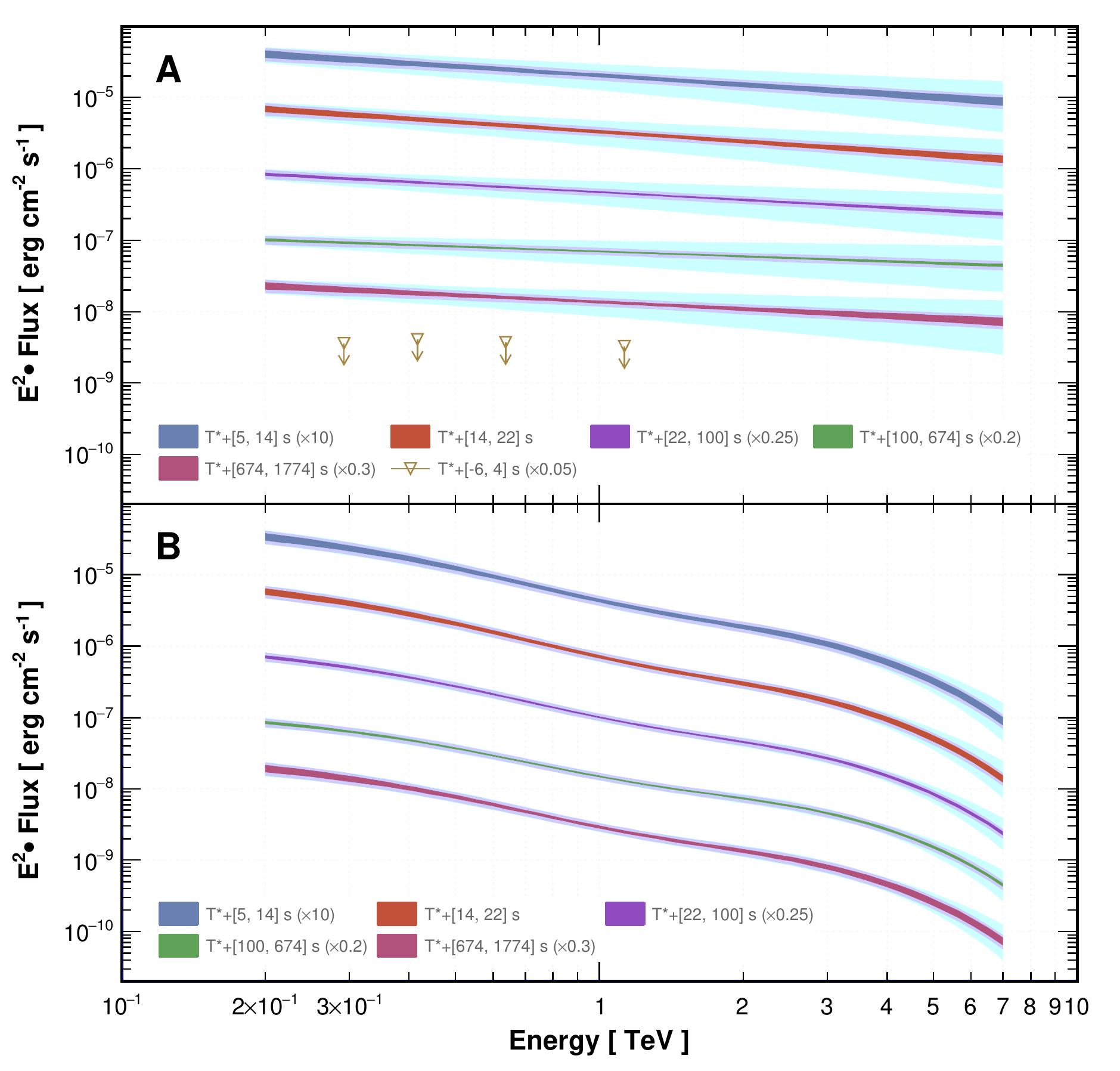}
\caption{{\bf Intrinsic and observed flux spectra for five time-intervals.}  (A) the intrinsic spectra corrected for EBL attenuation, assuming power-law functions  in the forward-folding calculation~\cite{Materials}; (B) the observed spectra obtained by re-applying the EBL attenuation to the corresponding intrinsic spectra. The colored bands indicate $1\,\sigma$ ranges of statistical uncertainties (the inner band) and systematic uncertainties (outer bands). The systematic uncertainties are further divided into detector-related (middle band in light blue) and EBL-related (outer band in light cyan) components. The 95\% upper limits on the intrinsic flux during the initial main burst phase ($T_0+[220,\,230]\,{\rm s}$) are indicated by dark-orange triangles in panel A. The EBL model~\cite{Saldana-Lopez2021} and its uncertainties are used to calculate the spectra and estimate the systematic uncertainties. The wave-like features in the observed spectrum are caused by the EBL attenuation at different wavelengths. An alternative calculation, independent on the EBL model, is shown in Fig.~\ref{fig:SED-comp}. Goodness of fit tests are shown in Fig.~\ref{fig:event_distribution} and Table~\ref{table:spectrum-events}.}
\label{fig:SED}
\end{figure}

Gamma rays emitted from distant astronomical sources are attenuated by photon interactions with the extragalactic background light (EBL). The gamma-ray spectrum that would be observed if the EBL were absent, referred to as the intrinsic spectrum, can be inferred from the observed events by correcting for EBL attenuation. We perform this correction by assuming an EBL model~\cite{Saldana-Lopez2021}. Fig.~\ref{fig:SED} presents both the observed and the EBL-corrected intrinsic flux spectra in the energy range from  $\sim 200$~GeV to  $\sim 7$~TeV, for five time-intervals during which the GRB was detected by LHAASO-WCDA. The time intervals were chosen to cover the  rising phase, the peak, and three periods in the decay phase. We fitted the intrinsic spectra with power-law model, which is consistent with the data up to at least 5~TeV  with no evidence for a spectral break or cut-off. The best fitting values of the spectral indices $\gamma$ of the power-law function are given in Table~\ref{table:spectrum-parameter}. These show a mild spectral hardening in time with $\Delta \gamma \sim 0.2$ between the first and last  time intervals. The uncertainty in  the EBL model~\cite{Saldana-Lopez2021}  is equivalent to a similar change in the spectral index (Table~\ref{table:spectrum-parameter}).
During  the initial main burst phase ($T_0+[220,\,230]\,{\rm s}$),  no  TeV emission  is detected (significance $<2.3\sigma$), with the $95\%$ upper limits on the flux  shown in Fig.~\ref{fig:SED}.

\begin{figure}
\centering
\includegraphics[width=0.9\linewidth]{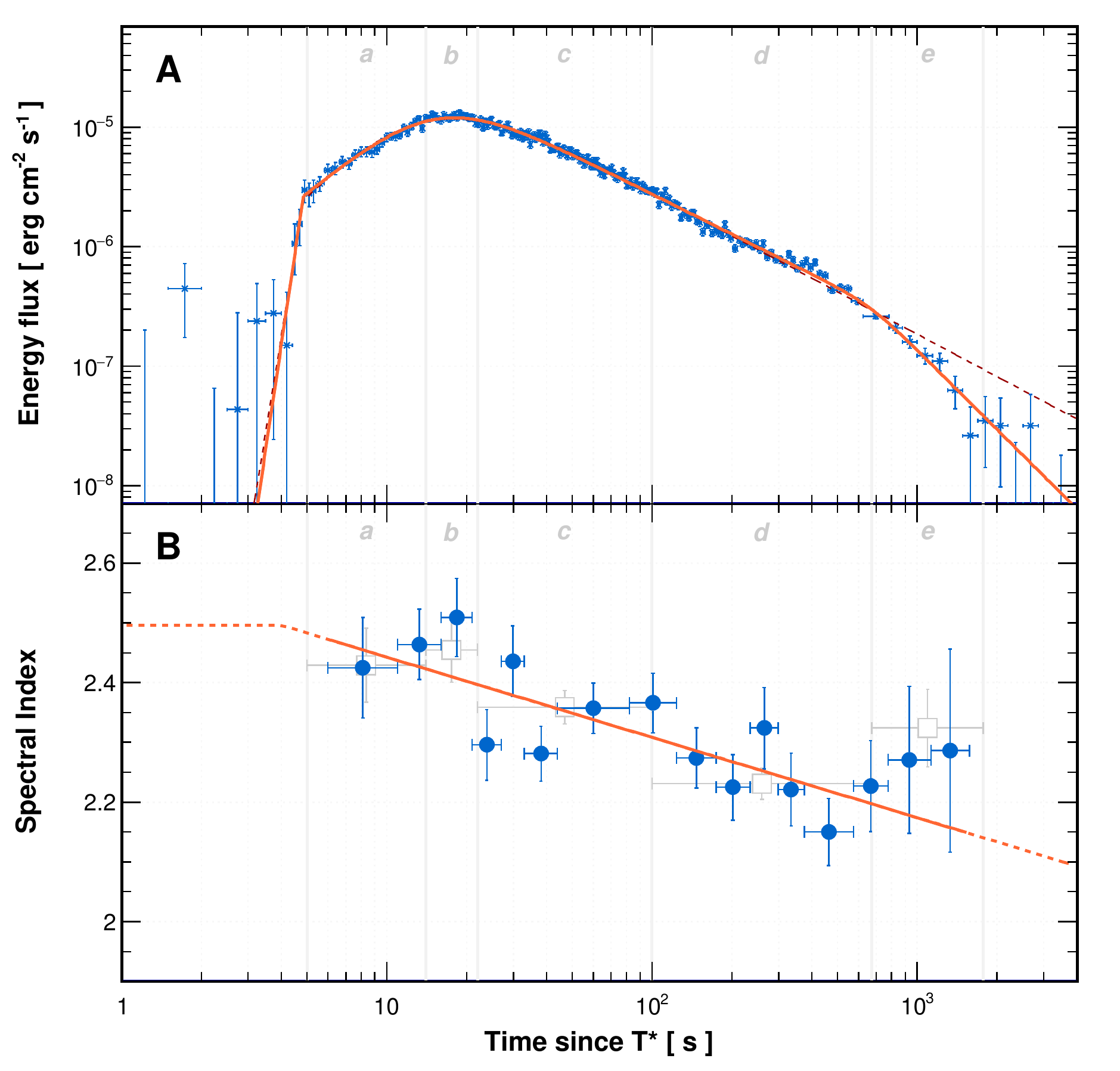}
\caption{{\bf Energy flux light curve and spectral evolution in the VHE band for GRB 221009A.} (A) the light curve in energy flux, integrated over the energy range 0.3 to 5~TeV. Blue points indicate the observations, with error bars indicating the $\sim 1\,\sigma$ statistical uncertainty (the system uncertainty $\sim 10.7\%$~\cite{Materials} is not included). The solid orange curve shows the fitted model, consisting of four joint power laws that describe the four-segment features in the light curve: rapid rise, slow rise, slow decay, and steep decay. The dark red dashed line shows the three-segment model, which has only one segment for the entire decay phase. The best-fitting parameter values for both models are listed in Table~\ref{table:LC-parameter}. (B) the temporal evolution of the power-law spectral index (blue circles) of photons, determined from the time-resolved intrinsic spectra. The orange line is the function $\gamma(t) = a\log(t) + b$ fitted to the data points. This model is flat before $T_0 + 230\,{\rm s}$. The time intervals for the spectrum fitting in Fig.~\ref{fig:SED} are labelled {\sl a}--{\sl e} in gray, in both panels. The average spectral indices of these five intervals are indicated by gray hollow squares in panel B.}
\label{fig:LC-log}
\end{figure}

The light curve of  of GRB 221009A converted to the energy flux,  integrated in the energy range 0.3--5~TeV, is presented in Fig.~\ref{fig:LC-log}. The peak observed flux is approximately $1.2\times 10^{-5}{\rm erg \, cm^{-2}\, s^{-1}}$, which corresponds to an apparent isotropic-equivalent luminosity of $L_{\rm iso, TeV} \sim 7.3\times 10^{50}\,{\rm erg\,s^{-1}}$ in  0.3--5~TeV, higher than other known sources at these energies.
Because  the Fermi GBM triggered on the precursor emission of this GRB, which was very weak compared to the main burst that started after a quiescent period of about 200~s (see Fig.~\ref{fig:rate_gbm_wcda}), we set   the reference time (hereafter $T^*$) of the afterglow light curves   at the  onset of the main component~\cite{Kobayashi&Zhang2007}.  Numerical studies have shown that this  is a  better approximation when investigating the afterglow emission~\cite{Kobayashi&Zhang2007,Lazzati2006}. 
The first main pulse of GRB 221009A lasted from 225~s to 228~s after the GBM trigger~\cite{GCN-HEBS,GCN-KW}, indicating that   $T^*$ is in the range of 225 to 228~s.  We measured $T^*$  by  fitting the LHAASO light curve with multi-segment power-laws, finding $T^*=225.7_{-3.2}^{+2.2}\,{\rm s}$ (Fig.~\ref{fig:LC-scanfit}). In the following analysis, we adopt $T^*=226\,{\rm s}$.

Fig.~\ref{fig:LC-log} has a  four-segment  shape, consisting of a rapid rise, a slow rise before the peak, a slow decay after the peak, and then a steep decay after a break, each of which is consistent with a power law function of time.   We fit  the data with a semi-smoothed-quadruple-power-law model (SSQPL, in which a smoothed triple power-law is  connected to an initial  rapid power-law rise)~\cite{Materials}. We find that the slope of the rapid rise is $\alpha_0 = 14.9_{-3.9}^{+5.7}$, the slope of the slow rise is $\alpha_1 = 1.82_{-0.18}^{+0.21}$,  the slope of the slow decay after the peak is  $\alpha_2 = -1.115_{-0.012}^{+0.012}$ and the slope of the steep decay after the break is $\alpha_3 = -2.21_{-0.83}^{+0.30}$. The transition time from the rapid rise to the slow rise phase is $T_{b,0} =T^*+ 4.85_{-0.10}^{+0.15}\,{\rm s}$, the peak time is $T_{\rm peak}=T^*+ 18.0_{-1.2}^{+1.2}\,{\rm s}$ and the break time to the steep decay is $T_{\rm b,2}=T^*+ 670_{-110}^{+230}\,{\rm s}$ (see Table~\ref{table:LC-parameter}). 

While the rapid rise is definitely present in the light curve (Fig.~\ref{fig:test_rapidrise}), the observations have insufficient temporal resolution to constrain the functional form of such a rapid rise. We assume a power law because it is  consistent with the other three  segments of the light curve.

We verify that the steep decay is required by  comparing the four-segment model with a three-segment model with only one segment for the whole decay phase  (Fig.~\ref{fig:LC-log}). We find that the four-segment model improves the fit over the three-segment model with a significance $9.2\,\sigma$~\cite{Materials},  indicating  a separate steep decay phase is justified by the data.  

\begin{figure}
\centering
\includegraphics[width=0.95\linewidth]{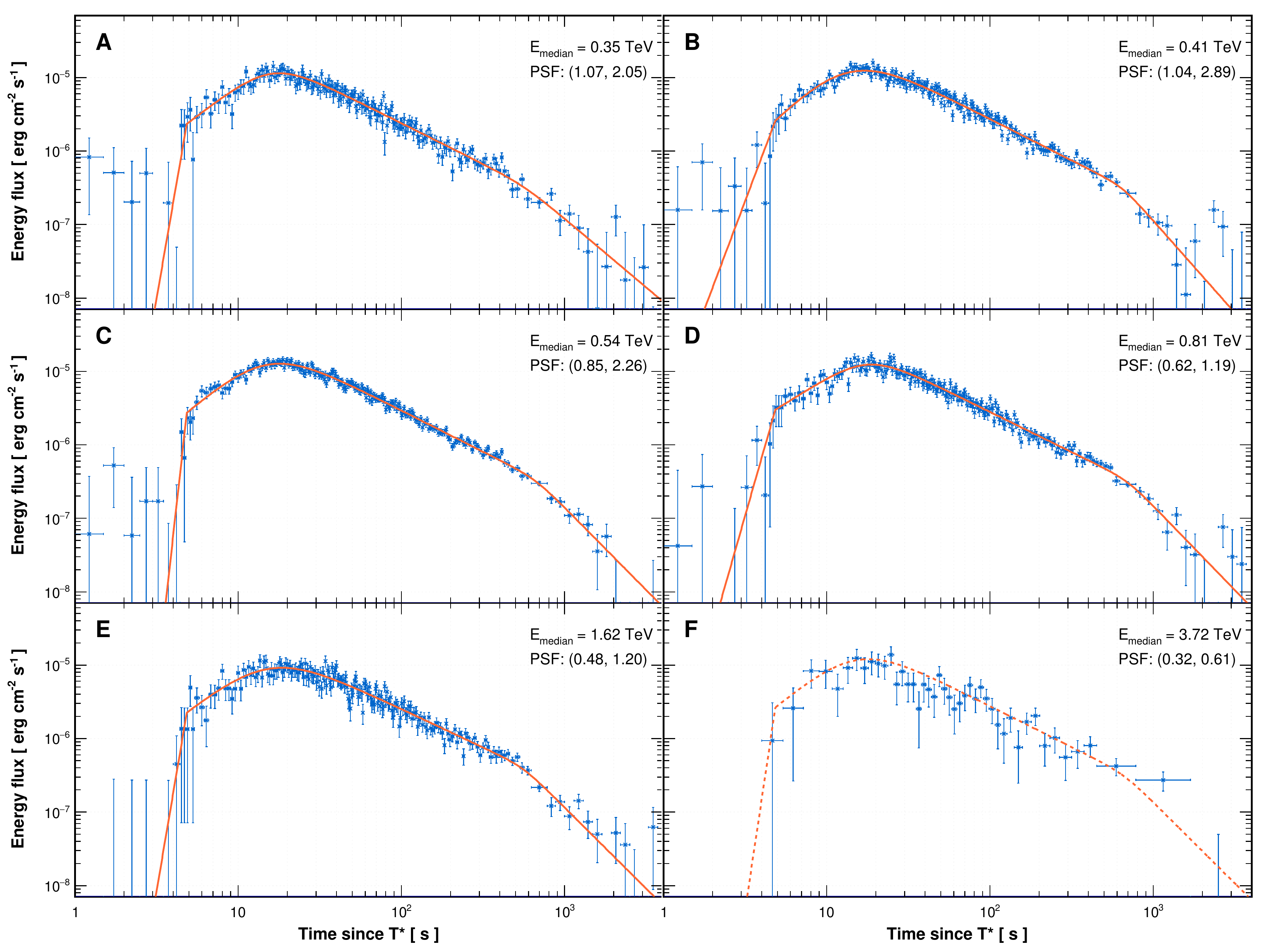}
\caption{{\bf Same as Figure~\ref{fig:LC-log}, but for six $\bm{N_{\rm hit}}$ segments.} The six segments are: $[30,\,33)$ (panel A), $[33,\,40)$ (panel B), $[40,\,63)$ (panel C), $[63,\,100)$ (panel D), $[100,\,250)$ (panel E) and $[250,\,+\infty)$ (panel F). The median energy ($E_{\rm median}$) and point spread function (PSF, 68\% and 99\% containment, in degrees) are labelled in each panel. The orange solid lines in panels A-E are four-segment models fitted to the data. The overall fit in Fig.~\ref{fig:LC-log} is shown as the dashed line in Panel F for comparison. During fitting, two parameters (the transition time from the rapid rise to the slow rise phase, and the sharpness of the transition from the slow decay to the steep decay phase) were fixed to the values obtained from Fig.~\ref{fig:LC-log}.}
\label{fig:LC-energies}
\end{figure}

We also apply the four-segment model to to light curves of sliced data samples with different energy bands (Fig.~\ref{fig:LC-energies}; and Fig.~\ref{fig:LC-comparison}). The best-fitting parameters are listed in Table~\ref{table:LC-parameter}. The peak times for different energy bands agree with each other, within the uncertainties, without any spectral evolution, indicating that the peak corresponds to the onset of the afterglow phase. The break in the decay phase of the light curve is observed in all energy bands. We find no systematic shift of the break time, which is also constant  within the uncertainties. 

We identify a small flare at $T^*+[320,\,550]\,{\rm s}$ (Fig.~\ref{fig:LC-mask}). To check if this flare affects the break at $t_{\rm b,2}$, we masked the data during the flare period when performing the fitting. This analysis shows that the break's behavior remained the same, indicating that the flare does not impact the identification of the steep decay.

A potentially  systematic uncertainty in  the  flux arises from the adopted EBL model. To test whether the uncertainty in the EBL  affects the light curve, we  re-calculate the light curve using the EBL intensities at the lower and upper boundaries of the EBL model~\cite{Saldana-Lopez2021}.  We then fitted the light curve with the same procedure  and the results are given in Table~\ref{table:LC-parameter}.
We find that the slopes and break times in the light curves remain almost unchanged, although the fluxes  change systematically. 

\section*{Interpretation of the TeV emission}
TeV gamma-ray emission could be produced by the  relativistic electrons accelerated by internal shocks~\cite{Bosnjak2009}  during the prompt emission  or external shocks during the afterglow phase~\cite{Sari&Esin2001,Wang2001,Zhang2001}.
The smooth temporal profile of the TeV emission in GRB 221009A suggests that it mainly results from an external shock.  
Because the synchrotron emission of relativistic electrons has a maximum energy $\ll 1$ TeV under the  assumption that the electrons are accelerated and radiate in the same zone, the TeV emission of GRB~221009A  is probably  produced by SSC  of relativistic electrons in the external shock, as has been proposed for previous TeV afterglows~\cite{190114C-2,Piran2019,Wang2019}. 
In the external shock model, the rise phase before the peak corresponds to the afterglow onset, where the forwardly-moving external shock sweeps up an increasing amount of ambient matter before being substantially decelerated (known as the coasting stage). The  density profile of the ambient matter can be described by $n(R)\propto R^{-k}$  ($R$ is the radius of the external shock), where $k=0$ corresponds to a homogeneous medium, while $k=2$ corresponds to a stellar wind from the GRB progenitor. 
In the  homogeneous medium case, the flux rises with time as $t^2$  during the coasting stage, where $t$ is the observer time since $T^*$, if the observed frequency is above the peak frequency of the SSC spectrum (i.e., in the spectral regime of $F_\nu \propto \nu^{-p/2}$, where $F_\nu$ is the flux density at the frequency $\nu$ and $p$ is the power-law index of the electron energy distribution)~\cite{Materials}. In the  stellar wind case, the light curve of the TeV afterglow is expected be flat or even declining with time in the same spectral regime~\cite{Materials,Fan2008}. To produce a rising flux in the wind medium case, the  Lorentz factor (defined as $\Gamma= 1/\sqrt{1-\beta^2}$, where $\beta$ is the velocity of the shock in unit of the speed of light $c$)  of   the forward shock must increase with time, perhaps due to energy injection. 
The rising slope  $\alpha_1 = 1.82_{-0.18}^{+0.21}$ is consistent with the homogeneous medium case without energy injection, so we infer $k=0$.
The initial rapid  rise phase has a high slope, albeit with a large uncertainty. $F_\nu\propto t^4$ might apply at this stage, because the spectrum is expected to be hard ($F_\nu \propto \nu^{-(p-1)/2}$) at such an early time~\cite{Materials}. This phase overlaps in time with the strongest pulses of the prompt main burst emission~\cite{GCN-HEBS,GCN-KW}, so energy into the external shock by the inner ejecta could take place, which may lead to a rapid flux increase~\cite{Materials}.

After the peak, the expected decay of the SSC emission is  $t^{-(9p-10)/8}$ in the spectral regime of $F_\nu \propto \nu^{-p/2}$  when the inverse-Compton scattering is in the Thomson regime~\cite{Wang2019}.  Depending on the parameter values, the scatterings could enter into Klein-Nishina (KN) regime, where the   photon energy, as measured in the rest frame of the upscattering electron, becomes comparable to the electron rest-mass energy.  In the KN regime, the decay is  $ t^{({\frac{3}{2}-\frac{5p}{4}})} $ in the spectral regime of $F_\nu \propto \nu^{-(p-1)}$~\cite{Materials}. Both interpretations are consistent with   the observed decay slope, $\alpha_2 = -1.115_{-0.012}^{+0.012}$, for  $p\sim 2.1$, although the observed spectrum in this period appears slightly softer than the  model prediction.

The light curve steepening at $t_{\rm b,2}\simeq 670\, {\rm s}$ after $T^*$ cannot be due to the KN scattering effect  because the spectrum after the break does not soften. 
The steepening resembles a jet break, which occurs when the  Lorentz factor  of a GRB jet drops to $1/\theta_0$, where $\theta_0$ is the initial half-opening angle of the jet. At this time, the  jet edge is visible to the observer, causing a steepening  in the light curve  by $t^{-3/4}$  a homogeneous medium~\cite{Meszaros&Rees1999,Dai2001a}.  If the lateral expansion of the jet becomes quick enough~\cite{Rhoads1999,Sari1999,Dai2001b}, a steeper decay is expected after the jet break for the VHE emission~\cite{Materials}. 
The  early jet break of GRB 221009A  implies a small  $\theta_0$, given by
\begin{equation}
\theta_0\sim 0.6^\circ E_{k,55}^{-1/8} n_{0}^{1/8} \left(\frac{t_{\rm b,2}}{670\,{\rm s}}\right)^{3/8},
\end{equation}
where $E_k$ is the isotropic kinetic energy of the ejecta and $n$ is number density of the circum-burst medium (where we adopt the convention that subscript numbers $x$ indicate normalisation by $10^x$ in cgs units).   This reduces the required energy in gamma-rays to  $E_{\gamma,j}=E_{\rm \gamma, iso}\theta_0^2/2\sim 5.5\times 10^{50} E_{\rm \gamma, iso,55}(\frac{\theta_0}{0.6^\circ})^2 \,{\rm erg}$ for GRB 221009A. This is consistent with the standard energy reservoir of GRB jets~\cite{Frail2001}. It has been suggested that GRBs could have a quasi-universal beaming configuration: a structured jet  with high anisotropy in its angular distribution of the fireball energy about the symmetry axis~\cite{Rossi2002,Zhang2002}. Under this assumption of a universal jet structure for GRBs, a  small opening angle of GRB 221009A could imply that  the brightest core of a structured jet was visible from Earth , explaining the high isotropic-equivalent energy of this GRB. Combined with the low-redshift of the burst, the small opening angle also explains the high fluence of this GRB.

Our identification of the TeV afterglow onset time can be used to estimate the initial bulk Lorentz factor  $\Gamma_0$ of the jet. The peak time ($t_{\rm peak}\sim 18 \, {\rm s}$ after $T^*$) of the light curve corresponds to the deceleration time, when most of the outflow energy is transferred to the shocked external medium. The initial bulk Lorentz factor is then
\begin{equation}
\Gamma_0=\left(\frac{3(1+z)^3E_k}{32\pi n m_p c^5 t_{\rm peak}^3}\right)^{1/8}=440 E_{k,55}^{1/8} n_{0}^{-1/8}\left(\frac{t_{\rm peak}}{{\rm 18\, s}}\right)^{-3/8},
\end{equation}
where  $m_p$ is the proton mass and $c$ is the speed of light. Because $\Gamma_0$ is almost insensitive to $E_k$ and $n$, the initial Lorentz factor of GRB~221009A is consistent with the upper range of values for previous GRBs with measured $\Gamma_0$ inferred from afterglow deceleration~\cite{Liang2010}. This implies that more energetic GRBs (in isotropic energy) have a larger initial Lorentz factor for the outflow.

\begin{figure}
\centering
\includegraphics[width=0.48\linewidth]{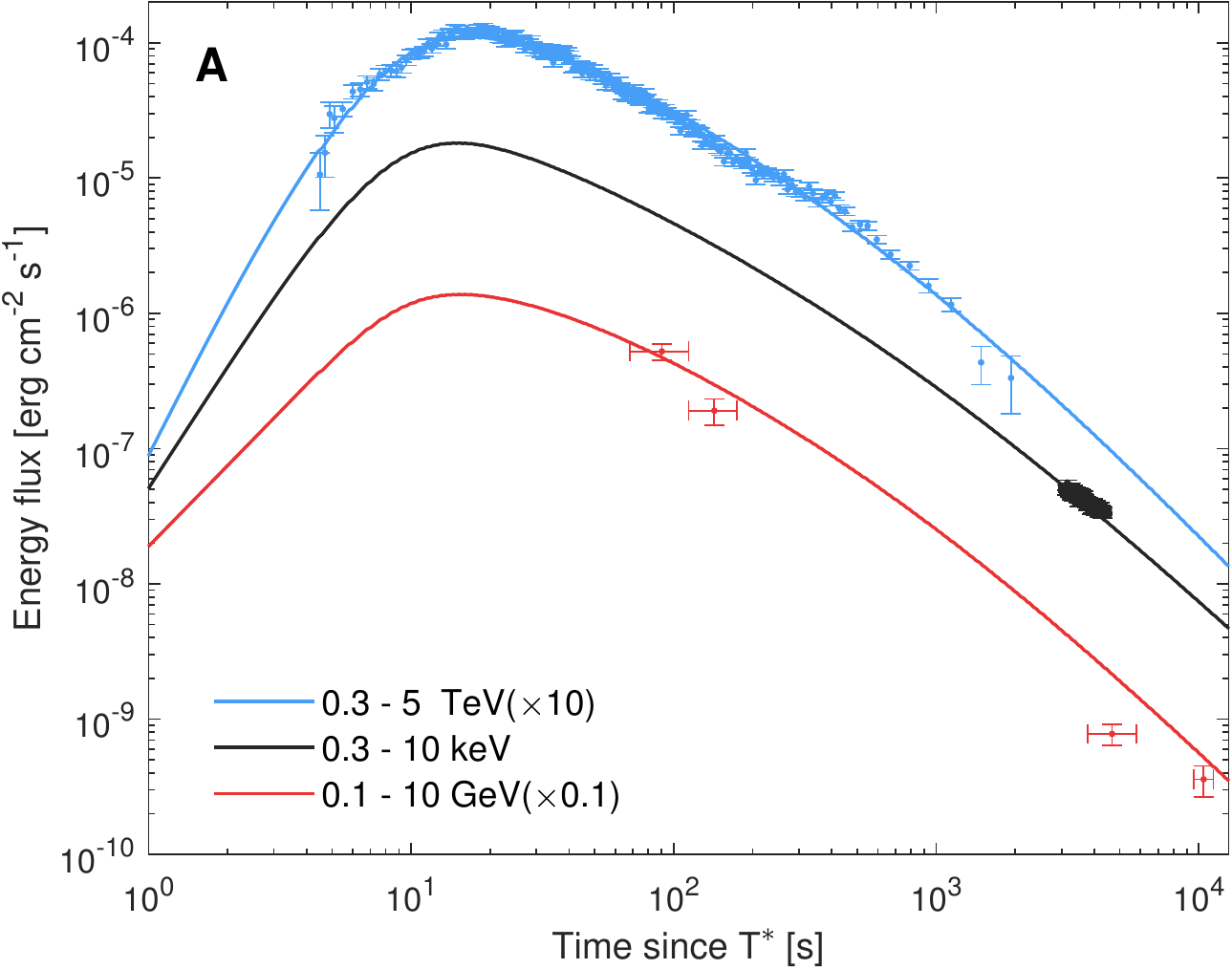}
\includegraphics[width=0.48\linewidth]{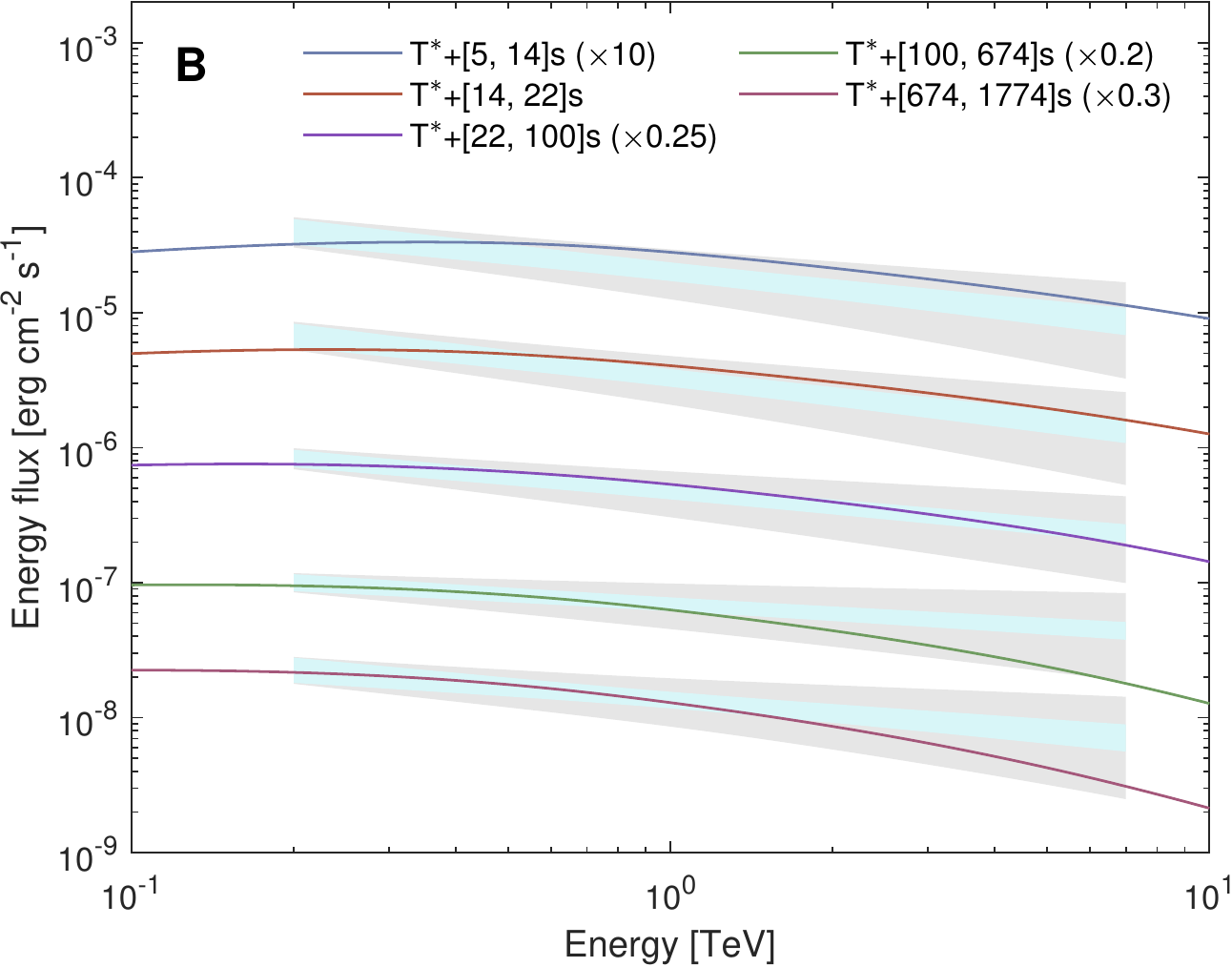}
\caption{{\bf Multi-wavelength modeling of GRB 221009A.} This model assumes afterglow emission arising from external forward shocks, emitting synchrotron and SSC radiation. (A)  the light curves in keV ($0.3$--$10\,{\rm keV}$, black), GeV ($0.1$--$10\,{\rm GeV}$, red) and TeV ($0.3$--$5\, {\rm TeV}$, blue) bands  for  the first $\sim 10^4\,{\rm s}$ after the burst.  Data points indicate the observations and curves are the output of the model.  (B) modeling of the LHAASO-WCDA spectra at five  time-intervals. The shade regions indicates $1 \sigma$ ranges of statistical uncertainties (inner band)
and systematic uncertainties (outer bands). The five colored lines indicate the output of the model for the five time-intervals. The  adopted parameters are: { $E_{k}= 1.5\times10^{55}\,{\rm erg}$, $\Gamma_0= 560$, $\epsilon_e= 0.025$, $\epsilon_B= 6\times 10^{-4}$, $p= 2.2$,  $n= 0.4\, {\rm cm^{-3}}$ and  $\theta_0= 0.8^\circ$ }. }
\label{fig:modeling}
\end{figure}
We performed  multi-wavelength modeling~\cite{Materials} of the Swift-XRT, Fermi-LAT~\cite{Liu2022} and LHAASO data assuming  synchrotron plus SSC radiation, within the framework of the afterglow
emission from external forward shocks. We use the   full Klein-Nishina cross section for the inverse Compton scattering 
and incorporate two-photon annihilation ($\gamma\gamma$) absorption within the source. For the jet break, we consider only the geometric effect when the jet edge is seen by the observer.  Because the inner core of the structured jet could be responsible for the early time afterglow emission, we consider only the data for  the first $\sim 10^4\,{\rm s}$ after the burst. The late-time afterglow emission could, in principle, include additional contributions from the outer wider components of the structured jet~\cite{Peng2005,Sato2021}.  
We find a model that is consistent with the  broadband light curves  and the LHAASO spectra  at various time intervals (Fig.~\ref{fig:modeling}),  under the following conditions~\cite{Materials}.
{ The initial isotropic-equivalent kinetic energy of the forward shock is $E_{k}\sim  1.5\times10^{55}\,{\rm erg}$ and the initial bulk Lorentz factor is $\Gamma_0\sim 560$.  The electrons and  magnetic
field behind the shock carry  fractions $\epsilon_e\sim 0.025$ and $\epsilon_B\sim 6\times 10^{-4}$, respectively, of the dissipated energy of the shock. The power law  index of the electron distribution is $p\sim 2.2$ and the density of the circum-burst medium is $n\sim 0.4\, {\rm cm^{-3}}$.} 
{Because there is degeneracy in the parameter space, the parameters are  not the only possible choice.} 
In this model, the X-ray  afterglow is produced by  synchrotron emission and the TeV afterglow is produced by  SSC emission, while the 0.1-10 GeV afterglow measured by Fermi-LAT has contributions from both synchrotron and SSC emission. 
We find that $\epsilon_e>\epsilon_B$ for the external shock, which is a necessary condition for  efficient SSC radiation.
The internal $\gamma\gamma$ absorption is  not strong, with  optical depth  $\le 1$  for gamma-rays of $5\, {\rm TeV}$~\cite{Materials}.
The half-opening angle in the modeling is $\theta_0\sim 0.8^\circ$, consistent with the analytical estimate.  
The resulting model SSC spectrum is harder than the observed spectrum at the low energy during  the first two time intervals (Fig.~\ref{fig:modeling}B), similar to  the previously studied TeV spectrum of GRB 190829A~\cite{190829A}. This discrepancy could be due to additional contributions to the flux measured by LHAASO-WCDA by some other emission processes, such as external inverse-Compton emission.

\section*{Limit on the prompt TeV emission }
In the optically-thin synchrotron scenario for  prompt emission in GRBs~\cite{Meszaros1994,Uhm2014}, the SSC emission produced by the same population of relativistic electrons  generates GeV to TeV gamma-rays~\cite{Bosnjak2009}. Previous observations obtained only loose upper limits on the TeV flux during the prompt emission phase~\cite{HAWC-prompt}. GRB 221009A was observed by LHAASO during the  main burst phase, yielding a differential flux limit of $\sim 6\times 10^{-8}\,{\rm erg\,cm^{-2}\,s^{-1}}$ at $\sim 1$ TeV from $T_0+220\,\textrm{s}$ to $T_0+230\,\textrm{s}$ (Fig.~\ref{fig:SED}).
Compared to the averaged MeV flux during the same period, which is $ 3\times 10^{-3}\,{\rm erg\, cm^{-2}\, s^{-1}}$(in the 20 keV - 15 MeV range; we regard this as a conservative estimate due to the saturation effect on the gamma-ray detectors~\cite{GCN-KW}),  the flux ratio  between TeV  and  MeV emission  is $\bar{R}\equiv F_{\rm TeV}/F_{\rm MeV}\le   2\times10^{-5}$. This is a stronger constraint than previous observations. 

The internal $\gamma\gamma$ absorption  suppresses the TeV flux during the prompt emission, because the radius of the internal shock or dissipation is much smaller than that of the external shock. The internal dissipation radius can be estimated if we know the variability timescale of the prompt emission. Before the saturation of the GBM data, the shortest variability timescale of GRB 221009A was $t_v\sim 0.082\, {\rm s}$~\cite{Liu2022}, implying the internal dissipation occurs at distance $R_{\rm in}\sim 2\Gamma_0^2 c  t_v=10^{15}\,{\rm cm}\,\left(\Gamma_0/440\right)^2 \left(t_v/0.082\, {\rm s}\right)$.  
We estimate the optical depth for TeV emission  to be $\tau_{\gamma\gamma}\sim 190\left({R_{\rm in}}/{10^{15}\,{\rm cm}}\right)^{-1}$~\cite{Materials}, producing strong attenuation of TeV photons,  which could explain the very low flux ratio between TeV  and  MeV emission.

In summary,  LHAASO observed the bright GRB 221009A at the  epochs  covering both the prompt emission phase and the early afterglow in the TeV band, revealing the onset of afterglow emission in the TeV band. We identify a jet break in the light curve of GRB 221009A,  indicating that the opening angle of GRB 221009A is $\sim 0.8^\circ$. Under the assumption of a universal jet structure for GRBs, this implies that the orientation of this GRB was such that the brightest core of a structured jet was visible from Earth, explaining the  brightness of this GRB.

\section*{References and Notes}
\newcommand{\etal}{\textit{et al.\/}}

\section*{Acknowledgments}
The LHAASO observatory, including its detector system, was designed and constructed by the LHAASO project team. Since its completion, it has been maintained by the LHAASO operating team. We extend our gratitude to all members of these two teams, especially those who work year-round at the LHAASO site, located over 4400 meters above sea level. Their tireless efforts ensure the detector and all its components, as well as the electricity power supply, operate smoothly.

\subsection*{Funding} 
This work was supported in China by the National Key R\&D program of China under grants 2018YFA0404201, 2018YFA0404202, 2018YFA0404203, and 2018YFA0404204, and by NSFC under grants U1831208, 11833003, 12022502, 12121003, U2031105, 12005246, 12173039, 1221101008, and 12203022. Support was provided by the National SKA Program of China under grant 2020SKA0120300, the Chinese Academy of Sciences under grant YSBR-061, the Department of Science and Technology of Sichuan Province under grant 2021YFSY0030, the Natural Science Foundation of Jiangsu Province under grant BK20220757, and the Chengdu Management Committee of Tianfu New Area for research with LHAASO data. In Thailand, support was provided by the National Science and Technology Development Agency (NSTDA) and the National Research Council of Thailand (NRCT) under the High-Potential Research Team Grant Program (N42A650868).

\subsection*{Author contributions}
Z.G.~Yao and X.Y.~Wang led the  data analysis and interpretation, respectively. S.C.~Hu (supervised by Z.G.~Yao) performed spectrum analysis. Y.~Huang (supervised by C.~Liu and Z.G.~Yao) calculated the light curve. H.C.~Li, C.~Liu, and Z.G.~Yao performed light curve fitting. M.~Zha coordinated the entire data analysis, provided reconstruction and simulation data, and participated in spectrum analysis. H.~Zhou and Zhen~Wang provided cross-checks.  Z.G.~Dai and B.~Zhang contributed to theoretical interpretation and manuscript structure. H.M.~Zhang performed multi-wavelength data analysis and contributed to interpretation. J.H.~Zheng (supervised by X.Y.~Wang) modeled the multi-wavelength data, and R.Y.~Liu contributed to the modeling of multi-wavelength data. Zhen~Cao is the spokesperson of the LHAASO Collaboration and the principal investigator of the LHAASO project, and coordinated the  working group for this study along with the corresponding authors. Internal reviews were provided by the Physics Coordination Committee of the LHAASO Collaboration led by S.Z.~Chen and the Publication Committee of the LHAASO Collaboration led by S.M.~Liu and D.~della~Volpe. All other authors participated in data analysis, including detector calibration, data processing, event reconstruction, data quality check,  simulations, and provided comments on the manuscript.

\subsection*{Competing interests}
There are no competing interests to declare.

\subsection*{Data and materials availability}

Data and software to reproduce our results are available at \url{https://www.nhepsdc.cn/resource/astro/lhaaso/paper.Science2023.adg9328/}. This includes the observed data and detector acceptance parameters used as input for our analysis, the resulting light curve data points (shown in Figs.~\ref{fig:LC-linear}, \ref{fig:LC-log}, \ref{fig:LC-energies}, \& \ref{fig:modeling}), spectrum data points (shown in Figs.~\ref{fig:SED} \& \ref{fig:SED-comp}), the code we used for the light curve fitting, and machine-readable versions of Tables~\ref{table:spectrum-events}, \ref{table:spectrum-parameter} \& \ref{table:LC-parameter}.


\clearpage


\setcounter{page}{1}
\renewcommand{\thepage}{S\arabic{page}}

\section*{}
\begin{center}
{\large Supplementary Materials for}

{\bf A tera-electronvolt afterglow  from a narrow jet in an extremely bright gamma-ray burst 221009A}

The LHAASO Collaboration$^\ast$\\

$^\ast$~Corresponding authors:  X.Y.~Wang (xywang@nju.edu.cn), Z.G.~Yao (yaozg@ihep.ac.cn), Z.G.~Dai (daizg@ustc.edu.cn), M.~Zha (zham@ihep.ac.cn), Y.~Huang (huangyong96@ihep.ac.cn),  J.H.~Zheng (mg21260020@smail.nju.edu.cn)\\
\end{center}

\vskip 2cm
\noindent{\bf This PDF file includes:}\\

\noindent\hspace*{1cm} LHAASO Collaboration Author List\\
\noindent\hspace*{1cm} Materials and Methods\\
\noindent\hspace*{1cm} Figures S1 to S10\\
\noindent\hspace*{1cm} Tables S1 to S3\\
\noindent\hspace*{1cm} References \textit{(45--58)}
\clearpage

\section*{LHAASO Collaboration authors and affiliations}

{
\noindent Zhen~Cao$^{1,2,3}$,
F.~Aharonian$^{4,5}$,
Q.~An$^{6,7}$,
Axikegu$^{8}$,
L.X.~Bai$^{9}$,
Y.X.~Bai$^{1,3}$,
Y.W.~Bao$^{10,33}$,
D.~Bastieri$^{11}$,
X.J.~Bi$^{1,2,3}$,
Y.J.~Bi$^{1,3}$
% ...
\\

\noindent \normalsize{$^{1}$Key Laboratory of Particle Astrophyics \& Experimental Physics Division \& Computing Center, Institute of High Energy Physics, Chinese Academy of Sciences, 100049 Beijing, China}\\
\normalsize{$^{2}$University of Chinese Academy of Sciences, 100049 Beijing, China}\\
\normalsize{$^{3}$Tianfu Cosmic Ray Research Center, 610000 Chengdu, Sichuan,  China}\\
\normalsize{$^{4}$Dublin Institute for Advanced Studies, 31 Fitzwilliam Place, 2 Dublin, Ireland }\\
\normalsize{$^{5}$Max-Planck-Institute for Nuclear Physics, P.O. Box 103980, 69029  Heidelberg, Germany}\\
\normalsize{$^{6}$State Key Laboratory of Particle Detection and Electronics, China}\\
\normalsize{$^{7}$University of Science and Technology of China, 230026 Hefei, Anhui, China}\\
\normalsize{$^{8}$School of Physical Science and Technology \&  School of Information Science and Technology, Southwest Jiaotong University, 610031 Chengdu, Sichuan, China}\\
\normalsize{$^{9}$College of Physics, Sichuan University, 610065 Chengdu, Sichuan, China}\\
\normalsize{$^{10}$School of Astronomy and Space Science, Nanjing University, 210023 Nanjing, Jiangsu, China}\\
\normalsize{$^{11}$Center for Astrophysics, Guangzhou University, 510006 Guangzhou, Guangdong, China}\\
\normalsize{$^{12}$Hebei Normal University, 050024 Shijiazhuang, Hebei, China}\\
\normalsize{$^{13}$Key Laboratory of Dark Matter and Space Astronomy \& Key Laboratory of Radio Astronomy, Purple Mountain Observatory, Chinese Academy of Sciences, 210023 Nanjing, Jiangsu, China}\\
\normalsize{$^{14}$Key Laboratory for Research in Galaxies and Cosmology, Shanghai Astronomical Observatory, Chinese Academy of Sciences, 200030 Shanghai, China}\\
\normalsize{$^{15}$Key Laboratory of Cosmic Rays (Tibet University), Ministry of Education, 850000 Lhasa, Tibet, China}\\
\normalsize{$^{16}$National Astronomical Observatories, Chinese Academy of Sciences, 100101 Beijing, China}\\
\normalsize{$^{17}$School of Physics and Astronomy (Zhuhai) \& School of Physics (Guangzhou) \& Sino-French
Institute of Nuclear Engineering and Technology (Zhuhai), Sun Yat-sen University, 519000 Zhuhai \& 510275 Guangzhou, Guangdong, China}\\
\normalsize{$^{18}$School of Physics and Astronomy, Yunnan University, 650091 Kunming, Yunnan, China}\\
\normalsize{$^{19}$D\'epartement de Physique Nucl\'eaire et Corpusculaire, Facult\'e de Sciences, Universit\'e de Gen\`eve, 24 Quai Ernest Ansermet, 1211 Geneva, Switzerland}\\
\normalsize{$^{20}$Institute of Frontier and Interdisciplinary Science, Shandong University, 266237 Qingdao, Shandong, China}\\
\normalsize{$^{21}$Department of Engineering Physics, Tsinghua University, 100084 Beijing, China}\\
\normalsize{$^{22}$School of Physics and Microelectronics, Zhengzhou University, 450001 Zhengzhou, Henan, China}\\
\normalsize{$^{23}$Yunnan Observatories, Chinese Academy of Sciences, 650216 Kunming, Yunnan, China}\\
\normalsize{$^{24}$Institute for Nuclear Research of Russian Academy of Sciences, 117312 Moscow, Russia}\\
\normalsize{$^{25}$Moscow Institute of Physics and Technology, 141700 Moscow, Russia}\\
\normalsize{$^{26}$Tsung-Dao Lee Institute \& School of Physics and Astronomy, Shanghai Jiao Tong University, 200240 Shanghai, China}\\
\normalsize{$^{27}$School of Physics, Peking University, 100871 Beijing, China}\\
\normalsize{$^{28}$School of Physical Science and Technology, Guangxi University, 530004 Nanning, Guangxi, China}\\
\normalsize{$^{29}$Department of Physics, Faculty of Science, Mahidol University, 10400 Bangkok, Thailand}\\
\normalsize{$^{30}$National Space Science Center, Chinese Academy of Sciences, 100190 Beijing, China}\\
\normalsize{$^{31}$Nevada Center for Astrophysics, University of Nevada, Las Vegas, NV 89154, USA}\\
\normalsize{$^{32}$Department of Physics and Astronomy, University of Nevada, Las Vegas, NV 89154, USA}\\
\normalsize{$^{33}$ Key laboratory of Modern Astronomy and Astrophysics (Nanjing University), Ministry of Education, Nanjing 210023, People Republic of China}\\
\noindent $\ast$~Corresponding authors:  X.Y.~Wang (xywang@nju.edu.cn), Z.G.~Yao (yaozg@ihep.ac.cn), Z.G.~Dai (daizg@ustc.edu.cn), M.~Zha (zham@ihep.ac.cn), Y.~Huang (huangyong96@ihep.ac.cn),  J.H.~Zheng (mg21260020@smail.nju.edu.cn)
}

\setcounter{figure}{0}
\setcounter{table}{0}
\renewcommand{\figurename}{Figure}
\renewcommand{\thefigure}{S\arabic{figure}} 
\renewcommand{\thetable}{S\arabic{table}}   
\setcounter{section}{0}
\renewcommand{\thesection}{S\arabic{section}} 
\setcounter{equation}{0}
\renewcommand{\theequation}{S\arabic{equation}} 

\section{Materials and Methods}

\subsection{LHAASO-WCDA \& the observation details} 
\begin{figure}[b]
\centering
\includegraphics[width=0.9\linewidth]{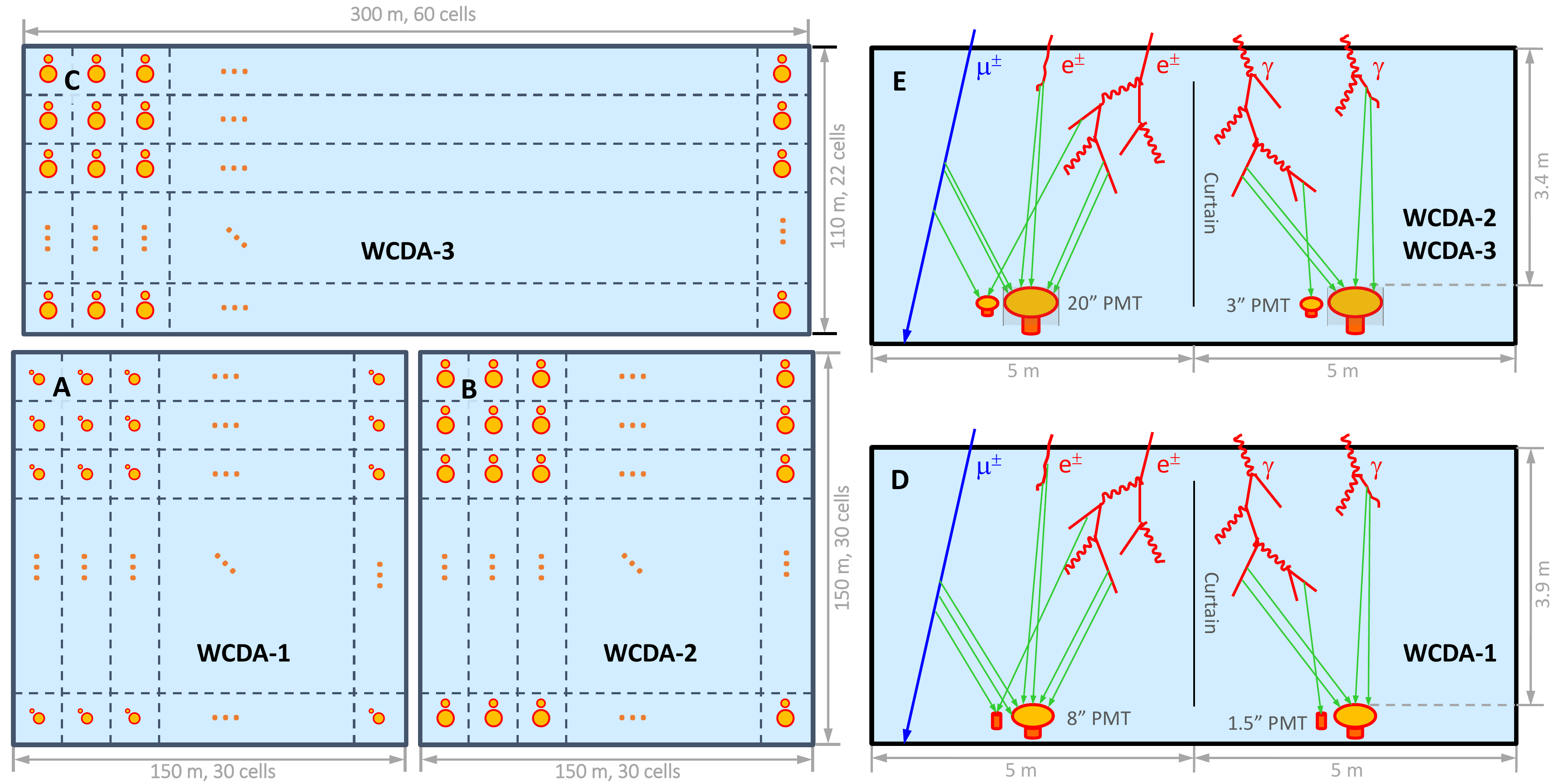}
\caption{{\bf Layout of LHAASO-WCDA.} (A--C) the detector layout for the three ponds, with black dashed lines representing the plastic curtains partitioning the array into cells. Two sets of circles representing the big (8-inch for WCDA-1, 20-inch for WCDA-2 and WCDA-3) and small (1.5-inch for WCDA-1, 3-inch for WCDA-2 and WCDA-3) PMTs. (D--E) show side views of two cells, as well as schematic physics processes (red wavy line:  Gamma; red straight line: electron or positron; blue line: muon; green line: Cherenkov light). The water depth over the PMT top surface is 3.9~m for WCDA-1 and 3.4~m for WCDA-2 \& 3.}
\label{fig:WCDA-fig}
\end{figure}

LHAASO is located at Mt. Haizi (29$^\circ$ 21' 27.56''N, 100$^\circ$ 08' 19.66''E)  and the altitude of 4410~m above sea level in Daocheng, Sichuan province, China. LHAASO is a hybrid extensive air shower observatory, in which WCDA is one of the major instruments, located at the center of the LHAASO~\cite{LHAASO}. WCDA consists of 3 water ponds with total area of $78,000\,\textrm{m}^2$ (Fig.~\ref{fig:WCDA-fig}). Each of the two smaller ponds has an effective area of $150\,\textrm{m} \times 150\,\textrm{m}$, and the third has an area of $300\,\textrm{m} \times 110\,\textrm{m}$. Every pond is divided into $5\,\textrm{m} \times 5\,\textrm{m}$ cells, which are separated by black plastic curtains. The first pond (WCDA-1) is equipped with 900 pairs of 8-inch and 1.5-inch photomultiplier tubes (PMTs), the second (WCDA-2) with 900 pairs of 20-inch and 3-inch PMTs, and the third (WCDA-3) with 1320 pairs of same PMTs as WCDA-2. Those PMTs reside at the bottom of the pond, facing upward, to collect the Cherenkov lights generated by charged particles in the water (Fig.~\ref{fig:WCDA-fig}D--E). Simulations have shown that the energy threshold of WCDA is $<100$~GeV. A more complete description of the WCDA detectors, the calibration and the reconstruction procedure, has been published elsewhere~\cite{WCDA-CPC}.

\begin{figure}
\centering
\includegraphics[width=0.8\linewidth]{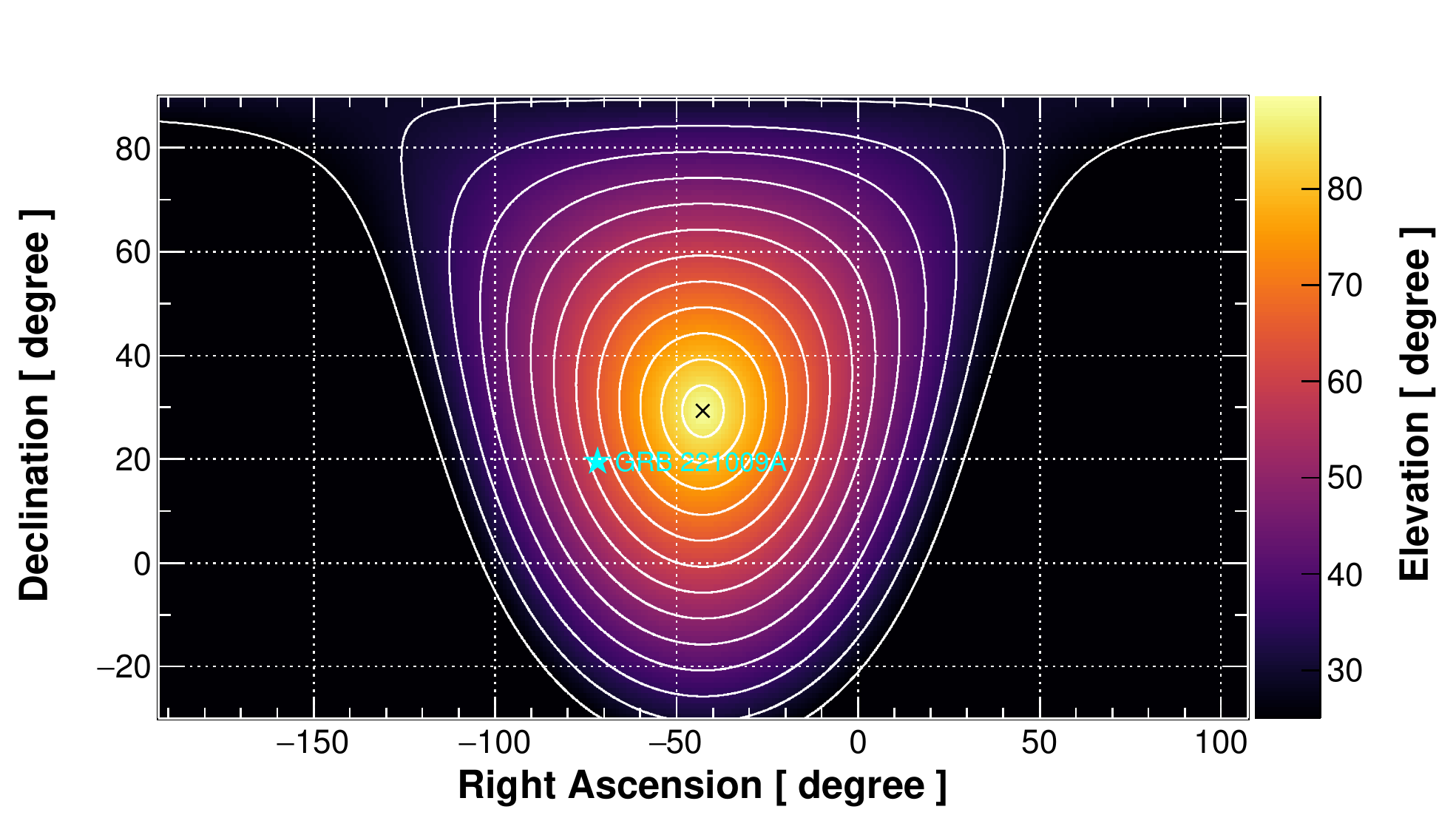}
\caption{{\bf GRB 221009A in the field-of-view of WCDA at the GBM trigger time $\bm{T_0}$.} 
The position of the GRB at $T_0$ is marked by a cyan star ($\star$), while the black cross ($\times$) represents the zenith of the view field at the same time. The colors indicate the elevations of the sky, with white contour lines every 5 degrees.}
\label{fig:wcda-fov}
\end{figure}

GRB 221009A occurred at within the WCDA field-of-view (FOV), as shown in Fig.~\ref{fig:wcda-fov}, where the FOV and GRB position at the GBM trigger time ($T_0$) is shown. The GRB has stayed in the FOV of WCDA until 10,000 seconds after $T_0$. The emission of the GRB became un-detectable in about 4000 seconds due to its rapid declining VHE flux and the lower efficiency of the detector at large zenith angles.

\begin{figure}
\centering
\includegraphics[width=0.8\linewidth]{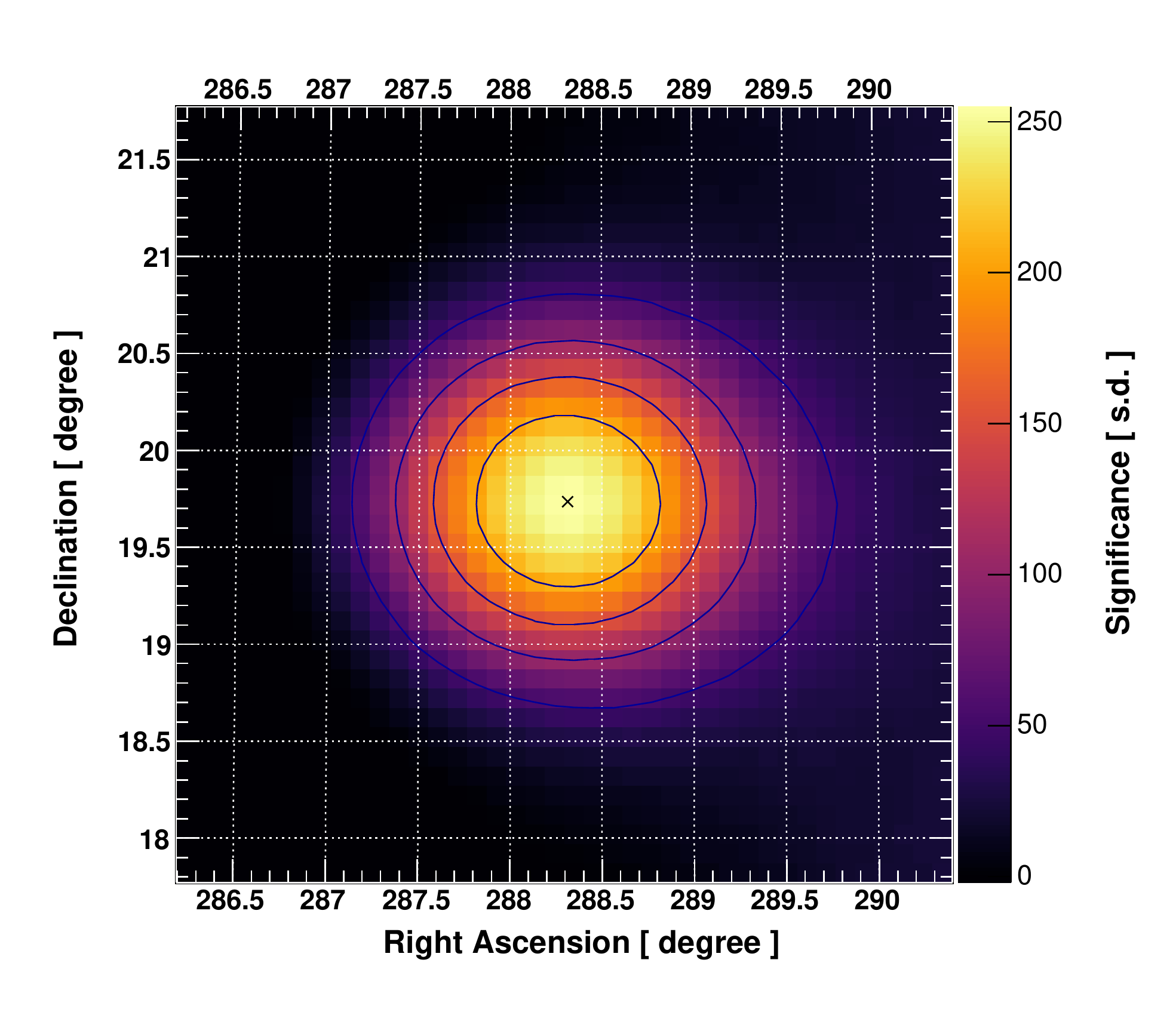}
\caption{{\bf Significance map of the GRB emissions detected by LHAASO-WCDA.} Reconstructed events with $N_{\rm hit}\geq 30$ as the baseline event selection criterion are used. The fitted position of the burst is indicated by a cross in the center of the map, with blue contour lines separated by 1/5 of the maximum significance.}
\label{fig:sigmap}
\end{figure}

The VHE emission from the GRB is bright, leading to a 1~kHz rise of the trigger rate in WCDA. 
We selected events with $N_{\rm hit} \geq 30$ within an angular radius of approximately $3.0^\circ$, where $N_{\rm hit}$ refers to the equivalent number of fired cells with a specific normalized charge threshold in a time window of $60\,{\rm ns}$ surrounding the reconstructed shower front. The cone with an aperture of $3.0^\circ$ contains over 99\% of events ($\phi_{68} = 0.83^\circ$ and $\phi_{99} = 2.29^\circ$).
The significance map of the signals from the GRB is obtained and is shown in Fig.~\ref{fig:sigmap}, where the backgrounds are estimated with four days' data around the burst time and in the same transit of the GRB, and the significance is evaluated with the Li-Ma formula~\cite{lima1983}.

\subsection{Spectrum analysis}

A forward-folding method~\cite{WCDA-CPC} is adopted to derive the spectra of the GRB in each time interval. The method uses a series of equations
\begin{equation}
N_{{\rm excess},i} = \int  {\rm d}E\int {\rm d}\theta\,\frac{{\rm d}t}{{\rm d}\theta} \phi(E) S_i(E,\theta)
\label{eq:forwardfold}
\end{equation}
to fit the parameters in the flux function $\phi(E)$, where $N_{{\rm excess},i}$ is the number of observed excess events in a particular segment $i$ that reflects the energy variations, $\theta$ is the zenith angle, ${\rm d}t$ is the time the source stayed in ${\rm d}\theta$, and $S_i(E,\theta)$ is the effective area of the detector at energy $E$ and zenith $\theta$. Effective area depends little on the azimuth, which are neglected. The parameter $N_{\rm hit}$, which is number of fired cells with a specific normalized charge threshold (equivalent to 0.5 photo-electron of the 8-inch PMT) in a time window of $60\,{\rm ns}$ surrounding the reconstructed shower front, is used to divide the events into 8 independent segments, $[30,\,40)$, $[40,\,63)$, $[63,\,100)$, $[100,\,160)$, $[160,\,250)$, $[250,\,400)$, $[400,\,500)$ and $[500,\,1000)$, designating different but overlapped energy ranges. The background is determined with data taken in the same transit as the GRB of two days before and after the burst, respectively. Specific event selection criteria were applied, including the Gamma-proton discrimination (relative spread of the lateral distribution of the charge, $P_{\rm n} < 1.2$), the location of the shower core inside the detector array (distance of core to the edge of the array, $D_{\rm c}>30\,{\rm m}$), and the reconstruction precision of the shower core (root of mean distance square of large hits to the shower core, $R_{\rm c}<10\,{\rm m}$), to improve the signal-to-noise ratio and to restrain the energy spread. For every $N_{\rm hit}$ segment, a particular angular range surrounding the GRB position, containing more than 99\% of events via fitting the spread of signals with a double-Gaussian function, is chosen as the source window. In the fitting, a power-law with an exponential cut-off $\phi(E) = \textrm{d}N/\textrm{d}E = A(E/{\rm TeV})^{-\gamma}\,e^{-E/E_{\rm cut}}$ is assumed as the function form of  observed spectrum, and a simple power-law $\textrm{d}N/\textrm{d}E = A(E/{\rm TeV})^{-\gamma}$ for the intrinsic one.  In the latter case, the photon attenuation derived from the EBL model~\cite{Saldana-Lopez2021} as a function of energy at the distance $z = 0.151$ is applied to correct the detector effective area, via re-weighting every simulated event with the survival probability ${\cal P}(E) = e^{-\tau(E)}$ from the EBL attenuation, where $\tau(E)$ is the optical depth of EBL at the photon energy $E$. This means that the effective area $S_i(E,\theta)$ in Eq.~\ref{eq:forwardfold} changes to $S_{{\rm int},i}(E,\theta) = e^{-\tau(E)}S_i(E,\theta)$ in calculating the intrinsic spectrum. The aforementioned method of calculating the observed spectrum is empirical and does not rely on EBL models. However, the observed spectrum can also be calculated by re-applying the EBL attenuation effect to the intrinsic spectrum. 

\begin{figure}
\centering
\includegraphics[width=0.8\linewidth]{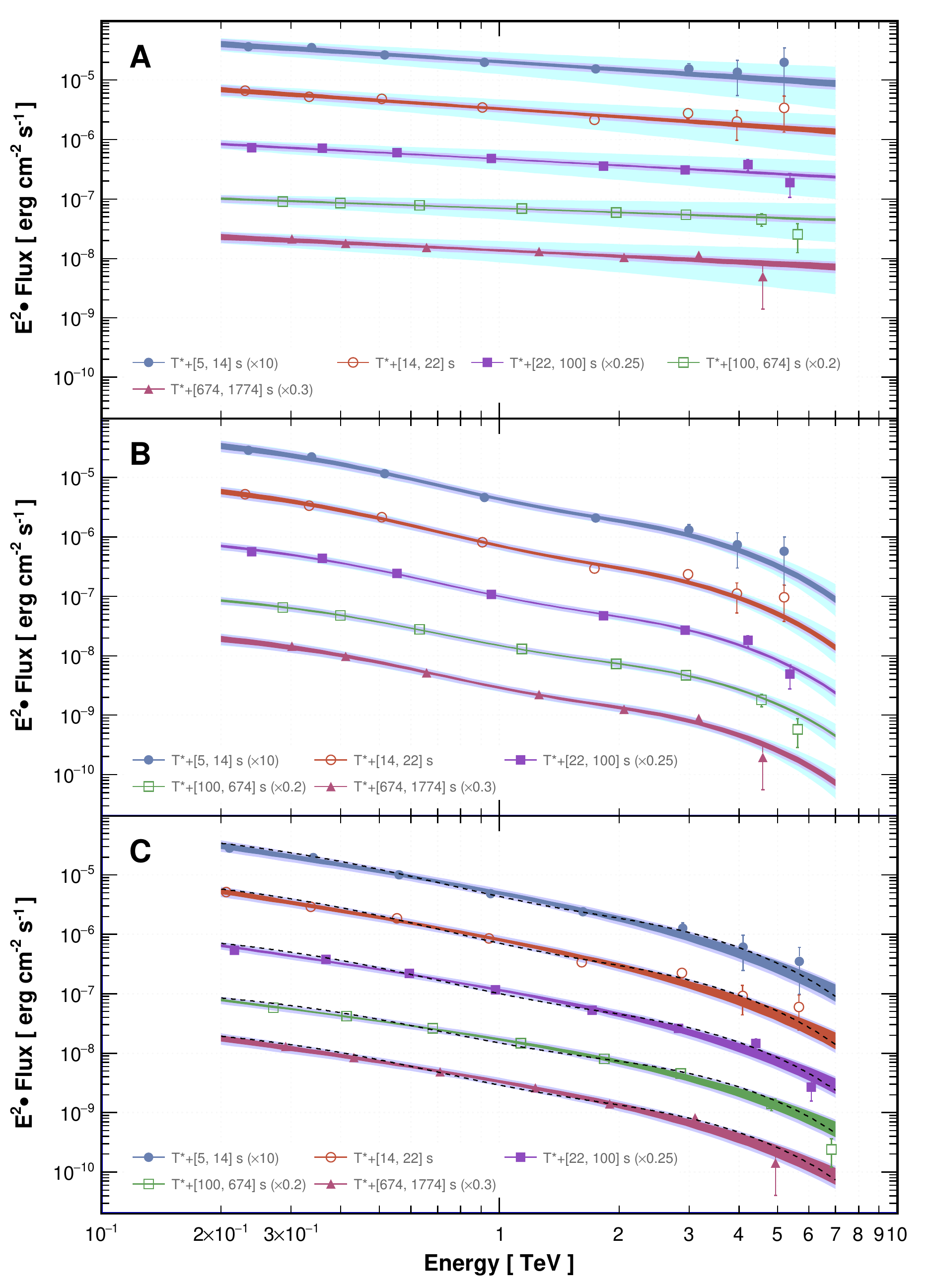}
\caption{{\bf Intrinsic and observed spectra with data points for five time intervals.} (A) the intrinsic spectra. (B) the observed spectra directly converted from the intrinsic ones. (C) the observed spectra obtained by assuming empirical power-law with exponential cut-off functions in which the EBL effect does not explicitly appear. Error bars for data points in all 3 panels indicate $1\,\sigma$ uncertainty. The spectral lines in Panel B are drawn as dashed black lines in Panel C for comparison. The meanings of color bands are the same as those in Fig.~\ref{fig:SED}.  The shifts in the energy values of data points between Panels B and C are due to high variations with energies in detector efficiency and poor energy resolution resulting from the use of $N_{\rm hit}$ to estimate energy. This introduces a dependence on the adopted function forms in forward-folding for determining median energies.}
\label{fig:SED-comp}
\end{figure}

\begin{figure}
\centering
\includegraphics[width=0.8\linewidth]{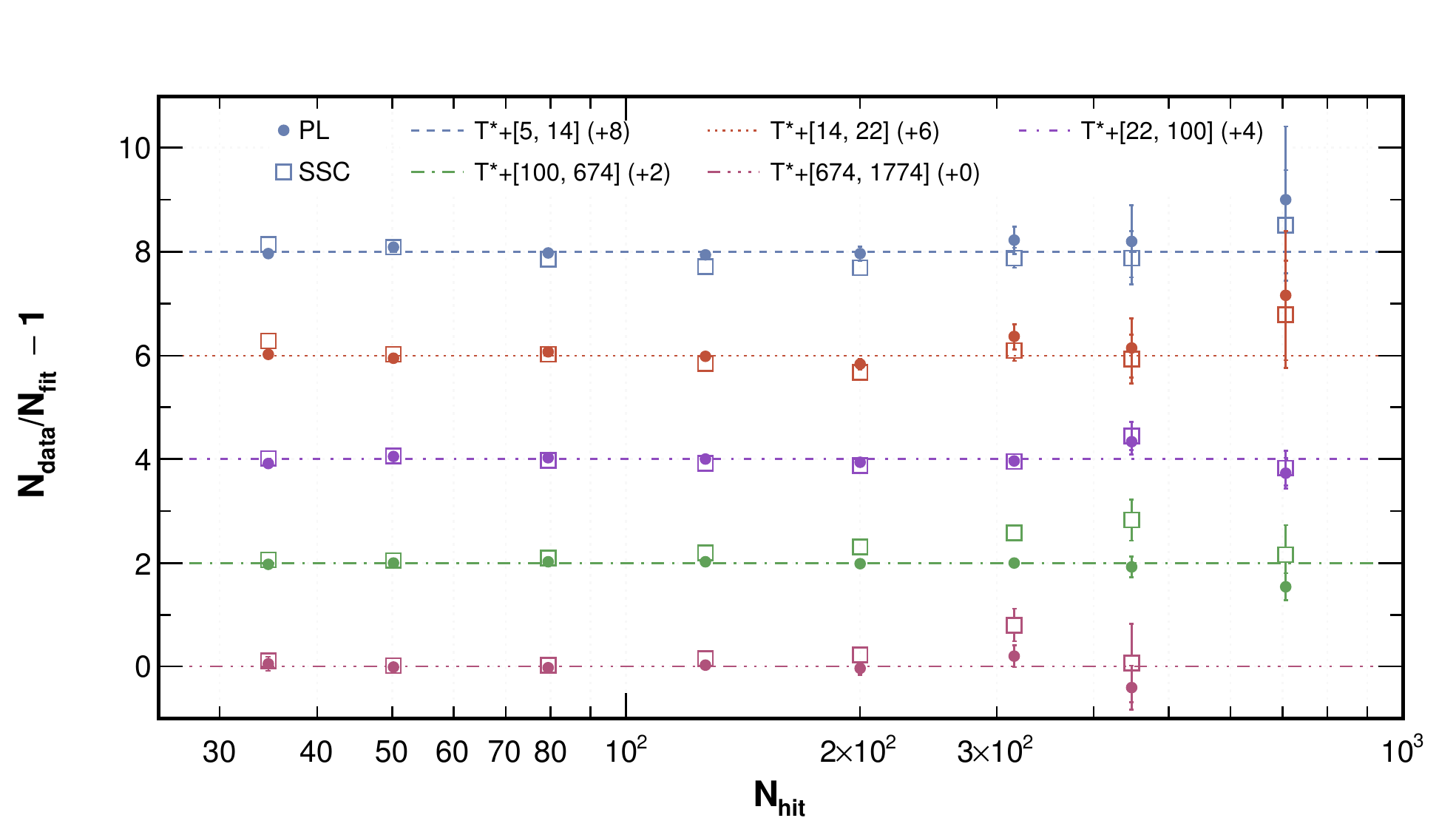}
\caption{{\bf Relative event distribution in every $\bm{N_{\rm hit}}$ bins for five time intervals.} The eight $N_{\rm hit}$ bins are defined as: $[30, 40)$, $[40, 63)$, $[63, 100)$, $[100, 160)$, $[160, 250)$, $[250, 400)$, $[400, 500)$, and $[500, 1000)$. The observed number of events is denoted by $N_{\rm data}$, while the number of expected events from either the intrinsic power-law fitting (solid circle) or the intrinsic SSC modeling (hollow square) is indicated by $N_{\rm fit}$. Data points have been shifted for display, with the corresponding values given in the legend. The data as well as error bars ($1\,\sigma$ uncertainty) used to generate this plot are in Table~\ref{table:spectrum-events}.}
\label{fig:event_distribution}
\end{figure}

After the spectrum is fitted, the data points are determined as follows. The energy is set to the median energy ($E_{\rm median}$) of the events expected to be observed by the detector in the corresponding $N_{\rm hit}$ bin, using the detector response with the fitted spectrum. The flux is then determined as
\begin{equation}
\phi_{\rm data} = \frac{N_{\rm data}}{N_{\rm fit}}\phi(E_{\rm median}).
\label{eq:phidata}
\end{equation}
Here, $N_{\rm fit}$ is the number of events calculated with the fitted spectrum $\phi(E)$ using the detector response in the $N_{\rm hit}$ bin, and $N_{\rm data}$ is the number of events that were actually observed in the $N_{\rm hit}$ bin. The spectra obtained with this procedure for the five time intervals are shown in Fig.~\ref{fig:SED-comp} and parameters are listed in Table~\ref{table:spectrum-parameter}. To test the reliability of the spectrum fitting, we provide the expected number of events from the intrinsic power-law function for each $N_{\rm hit}$ bin in Table~\ref{table:spectrum-events}, along with the corresponding number of observed events. Fig.~\ref{fig:event_distribution} displays a comparison between the fitted and observed event distributions in the $N_{\rm hit}$ bin domain.

To estimate the systematic uncertainty on the absolute flux, a comparison with the flux of the Crab Nebula is performed. The analysis uses 20 days of data from October 1 to 20, 2022, which reflect the detector's operating conditions during the observation of the GRB. The same event selection criteria applied in the GRB spectrum analysis are used to measure the Crab Nebula spectrum, which is then compared to the global fit function employing multi-wavelength observation data~\cite{LHAASO-Crab-paper}. The relative root-mean-square (RMS) difference between the measured and fitted spectra is evaluated in equal logarithmic energy intervals in the energy range of 0.3--5~TeV. The obtained RMS difference is 9.6\%. Due to poor Gamma-proton separation power, low detection efficiency at low energies, and low statistics at high energies, it is impractical to perform comparisons below 0.3~TeV or above 5~TeV in such a short period. Another approach to estimating the systematic uncertainty is to examine all the factors involved in the calculation of the detector response. While this is theoretically possible, it is difficult to achieve due to the complexity of the procedure and the presence of many unknown factors. Therefore, we prefer to use the aforementioned method of using the Crab Nebula as a standard candle, although it may underestimate the uncertainty (due to reliance on the global fit).

An additional systematic effect arises from uncertainties in the EBL model. To quantify the corresponding systematic uncertainties on the derived flux coefficients and power-law indices, we calculated intrinsic spectra corrected by two alternative EBL intensity models, which we refer to as low and high. These models correspond to the lower (weak attenuation) and upper (strong attenuation) boundaries of the uncertainty range in the model~\cite{Saldana-Lopez2021}. The obtained parameters are shown in Table~\ref{table:spectrum-parameter}. From these values, we find that the relative uncertainty of spectral indices is almost constant. For the low case, the spectrum becomes softer, and the relative difference in spectral indices is about -6.5\%. For the high case, the spectrum becomes harder, and the relative difference in spectral indices is about +6.5\%.

\subsection{Light curve calculation}

From $T_0 + 230\,{\rm s}$ to $T_0 + 2000\,{\rm s}$, intrinsic spectral indices for 15 time intervals (Fig.~\ref{fig:LC-log}) are calculated with the above-mentioned spectral analysis method. Those spectral indices are fitted with a function
\begin{equation}
\gamma(t) = a\log(t) + b
\label{eq:index_evolution}
\end{equation}
to obtain the time-resolved spectral index evolution, where $t = T - T^*$ is the time in second relative to $T^*$, while $T^* = T_0 + 226\,{\rm s}$ is chosen as the reference time (see main text). Since almost no signal is detected before $T_0 + 230\,{\rm s}$, we assume a constant spectral index that is connected to the fitting function at $t_{230} = T_0 + 230\,{\rm s} - T^*$, denoted as $\gamma(t) = \gamma(t_{230})$ when $t<t_{230}$. The parameters $a = -0.135 \pm 0.028$ and $b = 2.579 \pm 0.055$ are obtained by fitting the data, with a goodness-of-fit of $\chi^2/{\rm dof} = 16.1/14$, where dof is the degree-of-freedom. For comparison, a model with a constant spectral index is also used to perform the fit, which gives a spectral index of $2.320 \pm 0.015$ and a goodness-of-fit of $\chi^2/{\rm dof} = 39.7/15$. This is worse than the fit obtained using Eq.~\ref{eq:index_evolution}. Further analysis shows that using a constant spectral index would result in a 25\% shape distortion from the beginning to the end of the light curve, although it would not change the four-segment features of the light curve.

The WCDA detector's detection efficiency is calculated using a simulation data of the Crab Nebula. The efficiency is calculated for every simulated event on the Crab transit covering the GRB zenith in time range $T_0 + [200,\,4000]\,{\rm s}$, using several re-weighting factors that take into account the differences in transit time and intrinsic spectrum index between the Crab Nebula and the GRB. The energy range used in the simulation data is from 1~GeV to 1~PeV, which covers the range of the detector's response and the expected energies of GRB events. To estimate the background, data from the same transit as the GRB, two days before and after the burst, is used. No background rejection is applied in event selection to maximize the number of events, and the shower core selection is discarded as well. Six $N_{\rm hit}$ segments, $[30,\,33)$, $[33,\,40)$, $[40,\,63)$, $[63,\,100)$, $[100,\,250)$ and $[250,\,+\infty)$, are herein studied. Besides, overall events with $N_{\rm hit}\geq 30$ are particularly analyzed and focused, for the best statistical analysis of the light curve on the premise of a reliable energy spectrum measurement. In the analysis, the background and the detector efficiency coefficient for each $N_{\rm hit}$ segment, as a function of time, is fitted with cubic polynomials, in order to reduce fluctuations of the background and the simulation data. The cubic polynomials fit the data well, evidenced by reasonable $\chi^2/{\rm dof}$ values of the fittings, and no obvious transition of derivatives appeared in the fitting range. 

After obtaining the detector efficiency at every time slice of $0.1\,{\rm s}$, combining again the fitted intrinsic spectrum index $\gamma(t)$, and the number of excess events in the time slice, the flux coefficient $A(t)$ at 1~TeV were calculated. Integrating the energy in the range from $E_1 = 0.3\,{\rm TeV}$ to $E_2 = 5\,{\rm TeV}$,  which contains the main part of the GRB emissions detected by WCDA, and in which the intrinsic spectra is well measured for all time intervals as shown in Fig.~\ref{fig:SED}, the energy flux at the time $t = T-T^*$ is given by
\begin{equation}
\Phi(t) = \int_{E_1}^{E_2} E\cdot A(t)\, (E/{\rm TeV})^{-\gamma(t)}{\rm d}E.
\end{equation}
 Finally the light curve of energy flux as a function of time with a step size of $0.1\,{\rm s}$ is obtained.

The light curve of energy flux obtained from every $0.1\,{\rm s}$ is merged into wider bins to restrain fluctuations. The way of the binning for the case $N_{\rm hit} \geq 30$ is: binned with equal time $\Delta T = 0.5\,{\rm s}$ for $T \leq T_0 + 230\,{\rm s}$, in which no signals seem to be detected from the GRB; followed with a single time bin $T_0 + [230.0,\,230.4)\,{\rm s}$ as a transition; and then binned with equal $\Delta T = 0.2\,{\rm s}$ until $T  = T_0 + 245\,{\rm s}$ near the peak, for purpose of investigating the rise phase of the GRB emission in detail; after that, binned with equal relative errors at about 10\% level, up to around $T = T_0 + 1000\,{\rm s}$, in order to control the fluctuation consistently; afterwards, binned with equal $\Delta \log(t = T-T^*) = 0.057$ (where $T^* = T_0 + 226\,{\rm s}$), as wider and wider bins are required to deal with the weaker and weaker GRB signals. In each case, for every time bin, the geometric mean (square root) of the lower and upper edge is set as the timing point. The scenario of more sliced binning are studied as well, for instance doubling the total number of bins in the same time range, to check the consequences of different binning methods. No obvious influence to the light curve fitting is observed, as the binned Poisson likelihood method (discussed below) relies little on the binning, unless the binning is too coarse to reflect the features in the data to be fitted.

\subsection{Light curve fitting}
The light curve displays a four-segment feature, characterized by rapid rise, slow rise, slow decay, and steep decay phases. To fit this feature, we use a joint quadruple power-law function in which the slow rise and slow decay phase, and the slow decay and steep decay phase are smoothly connected using two sharpness parameters for the power-law breaks. However, we cannot apply the same approach to the transition from the rapid rise to the slow rise phase due to the insufficient number of excess events detected during the rapid rise phase. Therefore, we call the function a semi-smooth-quadruple-power-law (SSQPL) function, as follows.

The energy flux around the peak, which includes the slow rise and the slow decay phase, usually can be well described by a smoothly broken power-law (SBPL) function,
\begin{equation}
f_{12}(t) = A\left[\left(\frac{t}{t_{\rm b,1}}\right)^{-\omega_1\alpha_1} + \left(\frac{t}{t_{\rm b,1}}\right)^{-\omega_1\alpha_2}\right]^{-1/\omega_1},
\label{eq:f12}
\end{equation}
where $t = T - T^*$ is the time since the reference time $T^*$,  whose value is to be determined, along with other parameters, by fitting the data; $A$ is the energy flux coefficient; $\alpha_1$ and $\alpha_2$ are the power-law slopes before and after the break time $t_{\rm b,1}$ at which the slow rise transits into the slow decay; $\omega_1$ describes the sharpness of the break (a larger value corresponds to a sharper break). 
In case where another break occurs in the light curve, such as the break to the steep decay phase, a smooth triple power-law (STPL) model~\cite{Liang2008} can be used to fit the data, as follows,
\begin{equation}
f_{123}(t) =\left[f_{12}^{-\omega_2}(t) + f_3^{-\omega_2}(t) \right]^{-1/\omega_2},
\label{eq:f123}
\end{equation}
where $f_3(t)$ describes the shape of the steep decay phase:
\begin{equation}
f_3(t) = f_{12}(t_{\rm b,2})\left(t/t_{\rm b,2}\right)^{\alpha_3}.
\label{eq:f3}
\end{equation}
Here, $\alpha_3$ is the power-law index, and  $t_{\rm b,2}$ is the break time transiting into the steep decay with a sharpness $\omega_2$.
Lastly, we take into account the rapid rise phase, which can be formulated as
\begin{equation}
f_0 = f_{12}(t_{\rm b,0})\left(t/t_{\rm b,0}\right)^{\alpha_0}.
\label{eq:f0}
\end{equation}
Here, the parameter $t_{\rm b,0}$ represents the time when a break occurs between the rapid rise and slow rise phases, while $\alpha_0$ is the power-law index of the rapid rise phase. We assume that a sharp break occurs between these two phases, as the number of photon events acquired during the rapid rise phase is insufficient for obtaining the sharpness parameter through fitting. By combining $f_0$ and $f_{12}$, we obtain the joint function $f_{012}(t)$,
\begin{equation}
f_{012}(t) = \left\{
\begin{tabular}{ll}
$f_0(t)$ & $(t < t_{\rm b,0})$,\\
$f_{12}(t)$ & $(t \geq t_{\rm b,0})$.
\end{tabular}\right.
\label{eq:f012}
\end{equation}
Replacing $f_{12}(t)$ with $f_{012}(t)$ in Eq.~\ref{eq:f123}, the final so-called SSQPL function, which describes the entire four-segment light curve, is then obtained,
\begin{equation}
f_{0123}(t) =\left[f_{\rm 012}^{-\omega_2}(t) + f_3^{-\omega_2}(t) \right]^{-1/\omega_2}.
\label{eq:f0123}
\end{equation}
Altogether eleven parameters involved in this function are to be determined by fitting the data.

We use a binned Poisson likelihood method with the \textsc{MINUIT} package~\cite{minuit} to fit the data, and evaluate errors with \texttt{MINOS} in the same package. The method can be described as follows: For a particular time bin $i$, the number of observed events $N_{{\rm on}, i}$ is composed of the number of signals from the GRB $N_{{\rm s}, i}$ and the number of background events $N_{{\rm b}, i}$. This follows a Poisson distribution with a mean of $\mu_{{\rm b}, i} + \mu_{{\rm s}, i}$, where $\mu_{{\rm b}, i}$ is the expected number of background events and $\mu_{{\rm s}, i}$ is the expected number of signal events. We calculate $\mu_{{\rm s}, i}$ by converting the light curve function $f_{0123}(t)$ (Eq.~\ref{eq:f0123}) into the number of events in every 0.1~s time slice, and then summing over the time bin. The conversion of the light curve into the number of events is straightforward under the power-law assumption of the intrinsic spectrum, where the spectral index is known at any time from the evolution function. As mentioned in the previous section, the average number of background events for every time slice is obtained by a cubic polynomial fit to the background event distribution over time, so $\mu_{{\rm b}, i}$ can be treated as a known value with a negligible error. Therefore, we obtain the Poisson probability of the time bin when all parameters are set in $f_{0123}(t)$,
\begin{equation}
{{\cal P}_i} = \frac{e^{-(\mu_{{\rm b},i}+\mu_{{\rm s},i})}\,(\mu_{{\rm b},i}+\mu_{{\rm s},i})^{N_{{\rm on},i}}}{N_{{\rm on},i}!}.
\end{equation}
Multiplying ${\cal P}_i$ for all time bins within the fitting range from $T_1 = T_0 + 227\,{\rm s}$ to $T_2 \simeq T_0 + 4200\,{\rm s}$, we obtain the likelihood ${\cal L} = \prod_i {\cal P}i$. We construct $\lambda = -2\ln {\cal L}$ as a minimization evaluator to determine the values of parameters in $f_{0123}(t)$. To speed up the computation, we divide ${\cal L}$ by the likelihood ${\cal L}_0$ from a null hypothesis where no signal is detected from the GRB. This reduces the minimization evaluator to
\begin{equation}
\lambda_{\rm n} = -2\ln \frac{\cal L}{{\cal L}_0} =  \sum_i \mu_{{\rm s},i} + N_{{\rm on},i}\ln\frac{\mu_{{\rm b},i}}{\mu_{{\rm b},i}+\mu_{{\rm s},i}}.
\label{eq:lambdan}
\end{equation}

\begin{figure}
\centering
\includegraphics[width=0.8\linewidth]{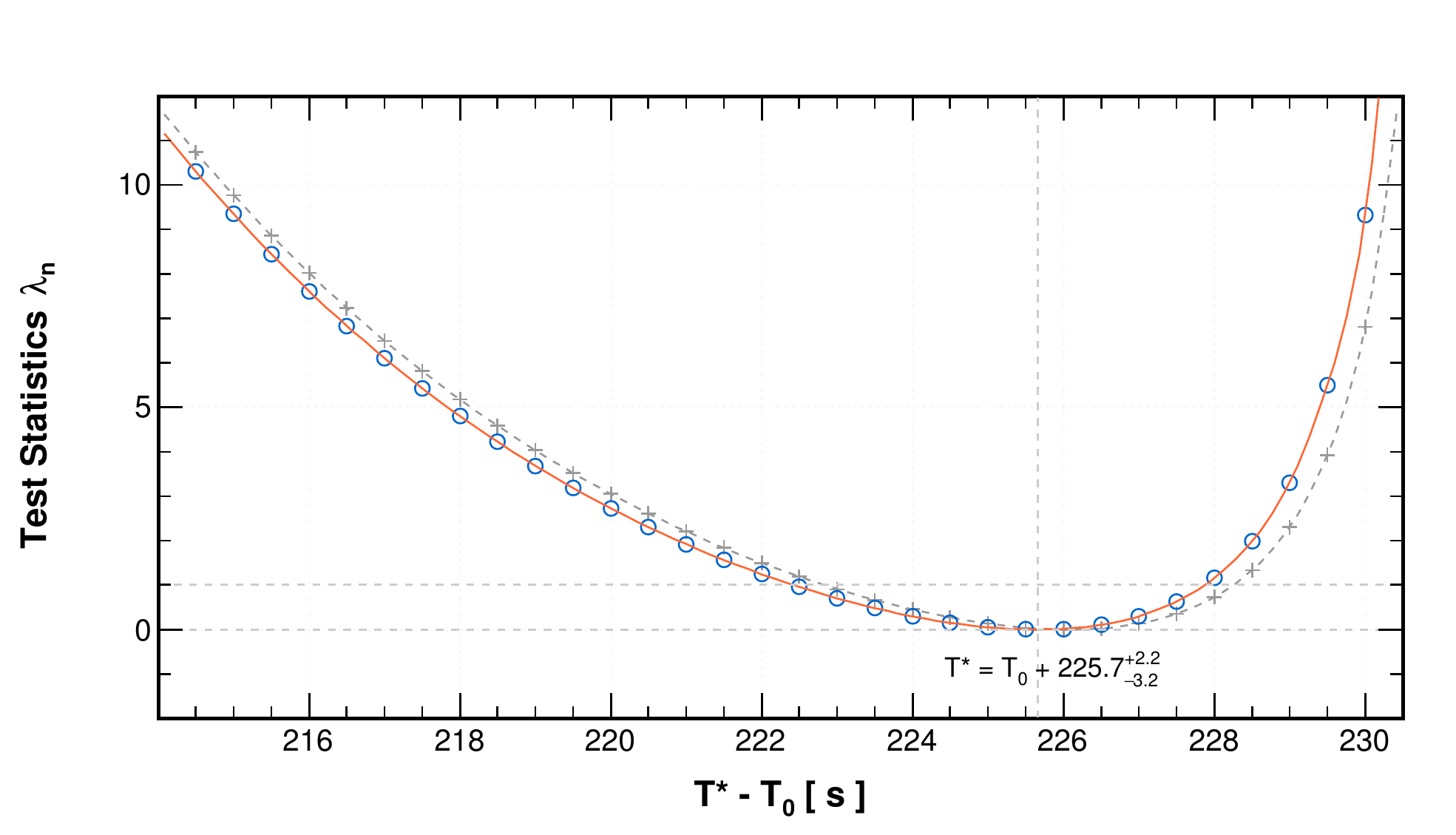}
\caption{{\bf Likelihood profile of $\bm{T^*}$ in the fitting of the afterglow light curve.} The reference time $T^*$ is fitted with a \texttt{SCAN} method. The test statistics, denoted as $\lambda_{\rm n}$ in Eq.~\ref{eq:lambdan}, is used to determine the minimum, which is set to zero in the plot. The orange solid curve represents the B-spline function that smoothly connects all scan points, resulting in a fitted value of $T^* = T_0 + 225.7_{-3.2}^{+2.2}\,{\rm s}$. To investigate the effect of ignoring the rapid rise phase, the lower threshold of the fitting range is increased from $T_1 = T_0 + 227\,{\rm s}$ to $T_1 = T_0 + 231\,{\rm s}$. The resulting gray crosses and dashed line indicate a fitted value of $T^* = T_0 + 226.1_{-3.2}^{+2.2}\,{\rm s}$.}
\label{fig:LC-scanfit}
\end{figure}

The gradient descent approach (\texttt{MIGRAD}), which is usually faster than other methods, is preferred in \textsc{MINUIT} for finding the minimum. The \texttt{SCAN} method is an alternative approach, which is slower, but can be well-controlled. A combination of the two methods is adopted in this analysis. Two parameters, the reference time $T^*$ of the burst and the sharp break time $T_{\rm b,0} = T^* + t_{\rm b,0}$ between the rapid rise and the slow rise phase, are manually scanned in turn for the best values within their reasonable ranges, while other parameters continue to be handled by \textsc{MINUIT} at every scan point. Several iterations are required to obtain the best values of both parameters simultaneously, even though there is very little correlation between the two parameters. Fig.~\ref{fig:LC-scanfit} displays the scan results of the $T^*$ parameters in the last iteration, while $T_{\rm b,0}$ is fixed to the best value in the last iteration, $T_0 + 230.85_{-0.10}^{+0.15}\,{\rm s}$. The obtained value of $T^* = T_0 + 225.7_{-3.2}^{+2.2}\,{\rm s}$ is close to $T_0 + 226\,{\rm s}$ and is consistent with the expectation from the analysis of the prompt emission profile. Therefore, we use $T^* = T_0 + 226\,{\rm s}$ for the analysis in this paper.

\begin{figure}
\centering
\includegraphics[width=0.8\linewidth]{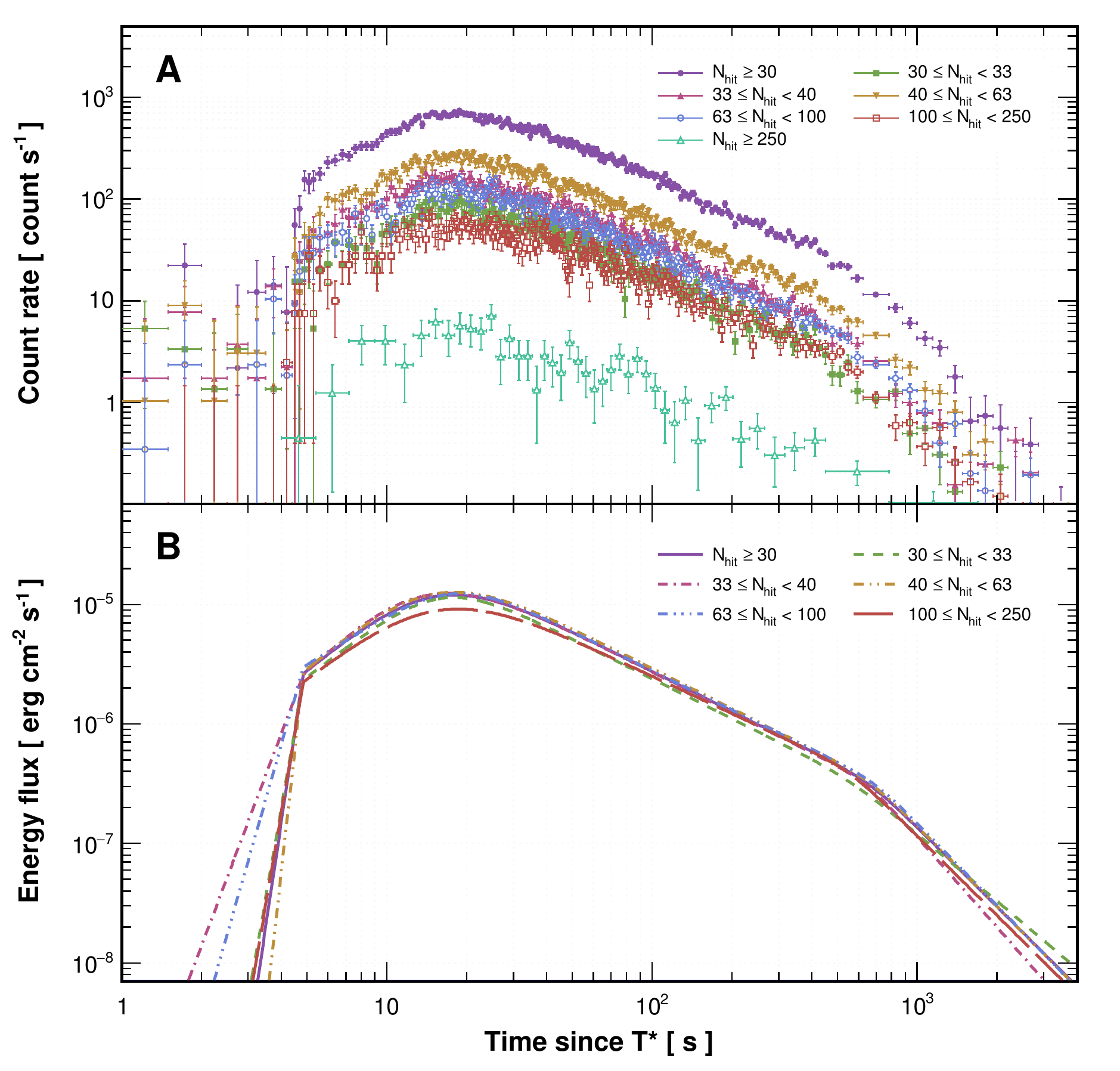}
\caption{{\bf Light curves of count rate and energy flux at different $\bm{N_{\rm hit}}$ segments.} (A) the count rate light curves, with error bars indicating $1\,\sigma$ uncertainty. (B) the energy flux light curves, which only includes the fitted curves (same as those shown in Fig.~\ref{fig:LC-log} and Fig.~\ref{fig:LC-energies}).}
\label{fig:LC-comparison}
\end{figure}

The final fitted parameters for various cases are listed in Table~\ref{table:LC-parameter}, including those from other $N_{\rm hit}$ segments, where parameters such as $T^*$ (reference time), $t_{\rm b,0}$ (break time from rapid rise to the slow rise phase), and $\omega_2$ (sharpness of the break between the slow decay and the steep decay phase) are fixed to the values obtained from the fitting in the case of $N_{\rm hit}\geq 30$. The same treatments are applied for other scenarios unless explicitly stated. It is worth mentioning that the light curves calculated from different $N_{\rm hit}$ segments appear nearly the same despite their different count rates, as shown in Fig.~\ref{fig:LC-comparison}. This is due to the power-law shape of the intrinsic spectra at all time intervals. 

To test the presence of the steep decay phase, i.e., a break in the decaying process after the peak at $T_{\rm b,2} = T^* + t_{\rm b,2} \simeq T_0 + 900\,{\rm s}$ (the best fitted value $T_{\rm b,2} = T_0 + 896_{-90}^{+230}\,{\rm s}$), we compare $f_{0123}(t)$ (Eq.~\ref{eq:f0123}) and $f_{012}(t)$ (Eq.~\ref{eq:f012}) in fitting the data. The results are shown in Fig.~\ref{fig:LC-log}A. The maximum likelihood value for the case with and without the break, denoted as ${\cal L}_{\rm max}$ and ${\cal L}_{\rm 0,max}$, is calculated, leading to the likelihood ratio $\chi^2({\rm dof} = 3) = -2\ln {\cal L}_{\rm 0,max}/{\cal L}_{\rm max} = 94.4$. This corresponds to a significance $9.2\,\sigma$, supporting the existence of the break.

\begin{figure}
\centering
\includegraphics[width=0.8\linewidth]{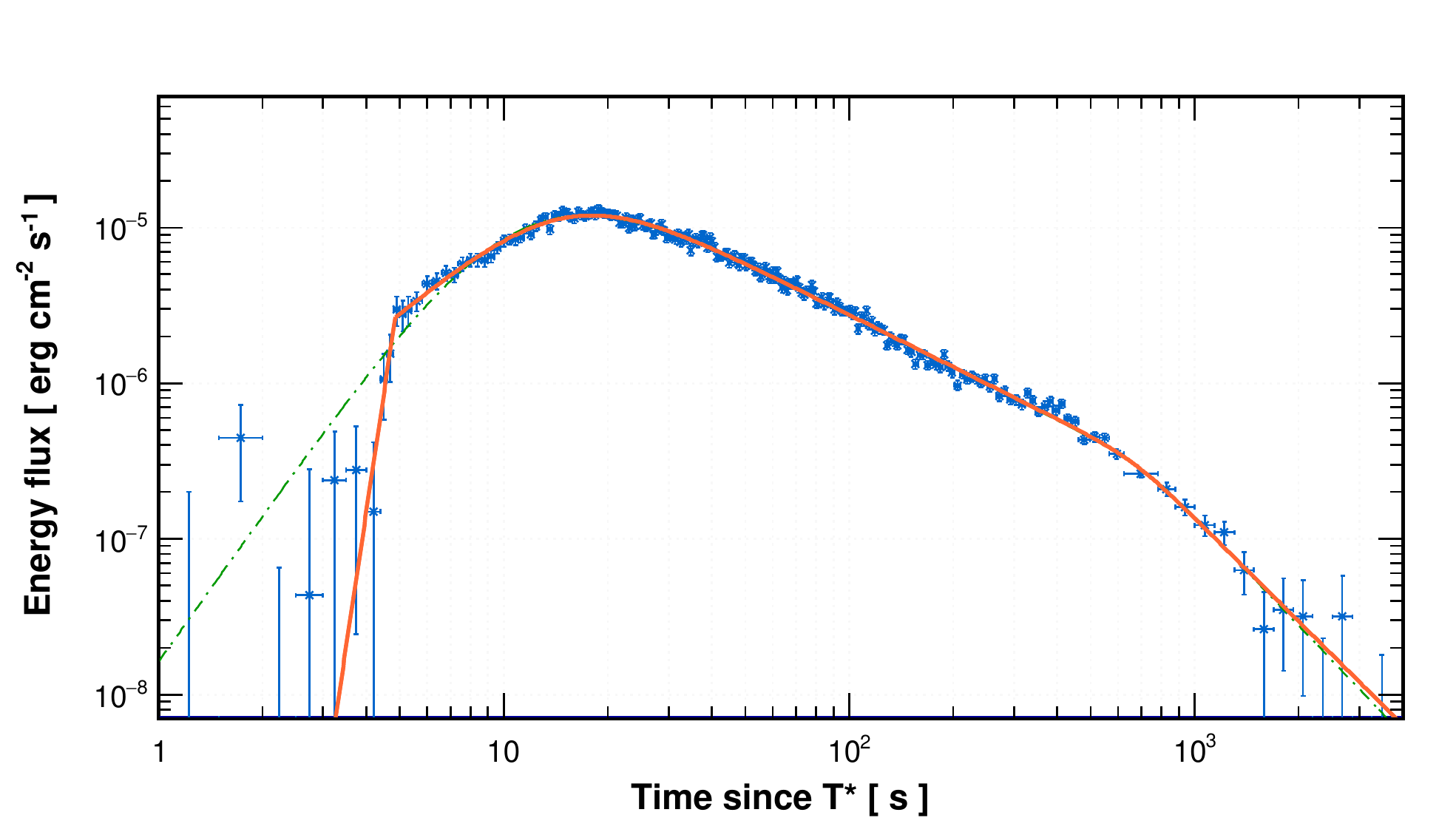}
\caption{{\bf Test of the presence of the rapid rise phase.}  The data points and the orange solid line are identical to those displayed in Fig.~\ref{fig:LC-log}, while the dash-dotted line in green represents the fitting function that assumes one power-law for the entire rising phase. The test shows that the significance of presence of the rapid rise is $5.9\,\sigma$.}
\label{fig:test_rapidrise}
\end{figure}

To test the presence of the rapid rise phase, i.e., a break in the rising process before the peak at $T_{\rm b,0} = T^* + t_{\rm b,0} \simeq T_0 + 231\,{\rm s}$ (the best fitted value $T_{\rm b,0} = 230.85_{-0.10}^{+0.15}\,{\rm s}$), we compare $f_{0123}(t)$ (Eq.~\ref{eq:f0123}) and $f_{123}(t)$ (Eq.~\ref{eq:f123}) in fitting the data. The results are shown in Fig.~\ref{fig:test_rapidrise}. The maximum likelihood value for the case with and without the break, denoted as ${\cal L}_{\rm max}$ and ${\cal L}_{\rm 0,max}$, is calculated, leading to the likelihood ratio $\chi^2({\rm dof} = 2) = -2\ln {\cal L}_{\rm 0,max}/{\cal L}_{\rm max} = 38.7$. This corresponds to a significance $5.9\,\sigma$, supporting the existence of the break. We consider whether the fit range influences the result of the hypothesis test. Iterating the cases of $T_1 = T_0 + {228,\,229,\,230}\,{\rm s}$ for the lower threshold of the fitting range to replace the original $T_1 = T_0 + 227\,{\rm s}$, the corresponding significance is 5.9, 5.8 and 5.6~$\sigma$, respectively, so little difference is found.

There appears to be a small flare during $T=T^*+[320,\,550]\,{\rm s}$ in the light curve. We test whether this flare affects the break by masking the data during the flare period in performing a similar fitting. No influence is found.

\begin{figure}
\centering
\includegraphics[width=0.8\linewidth]{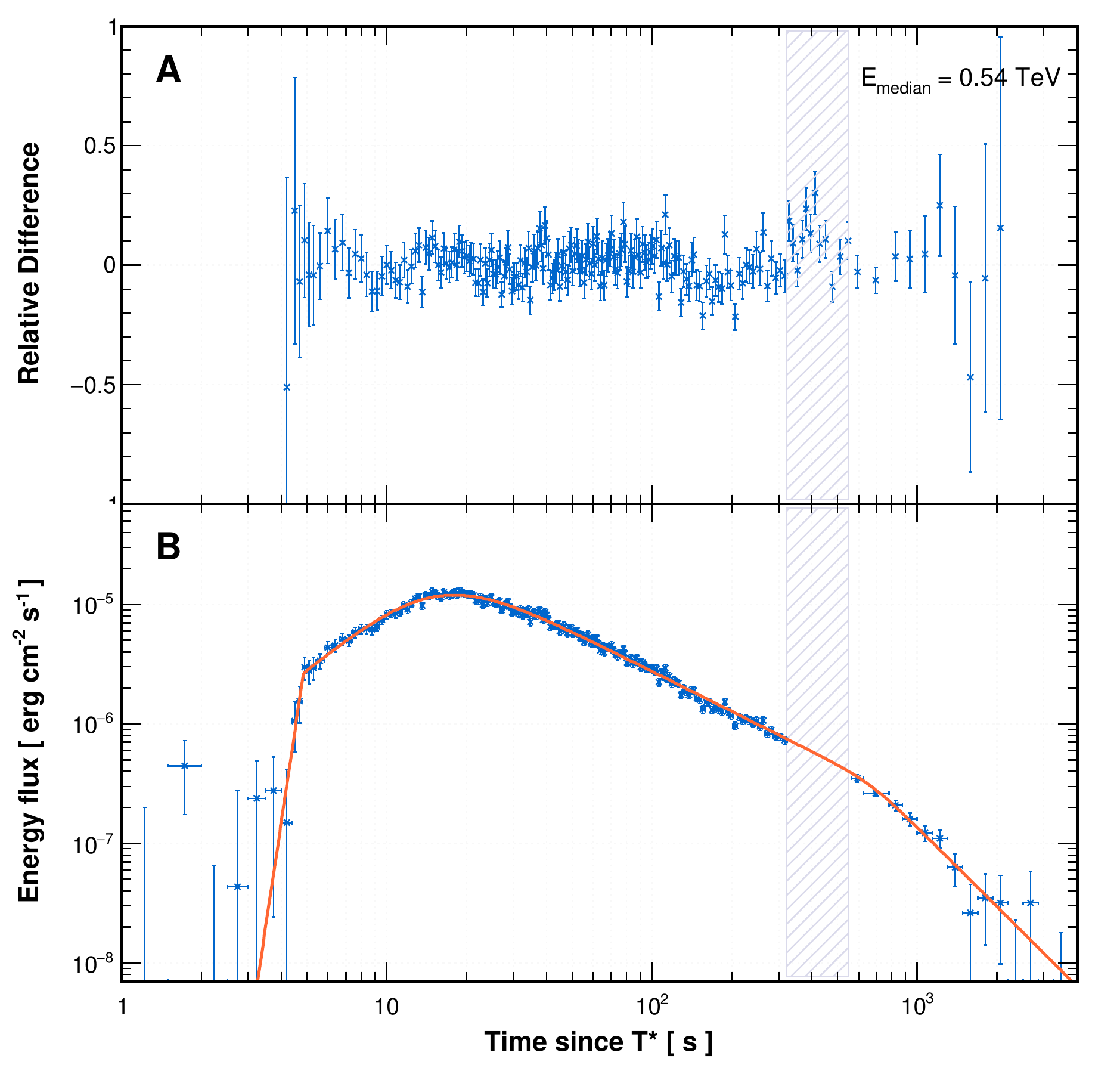}
\caption{{\bf Influence of the suspicious flare on the light curve fitting.} (A) relative difference ($\phi_{\rm data}(t)/\phi_{\rm fit}(t) - 1$) between the data and the fitting function. (B) light curves fitting for GRB 221009A after masking the flare period. Error bars in both panels indicate $1\,\sigma$ uncertainty. The shadow area indicates the masked period, which is obtained from the behaviour of relative differences of signals in Panel A.}
\label{fig:LC-mask}
\end{figure}

The systematic uncertainty on the light curve calculation can be attributed to three factors: flux calculation, EBL model, and index evolution model. To estimate the uncertainty due to the flux calculation, the observed GRB spectrum around the light curve peak is compared to the spectrum obtained with the criteria used for the spectrum measurement, and the relative RMS difference is found to be 4.8\% for the energy range of 0.2--5~TeV. Combining this with the previously obtained 9.6\% uncertainty due to the comparison of the Crab Nebula spectrum, the overall uncertainty for the energy flux calculation is estimated to be 10.7\%. To check the uncertainty due to the EBL model, the same analysis is performed using two additional EBL intensity models, corresponding to the lower and upper boundary of the error range in the model, resulting in a relative difference of about 27\% in the flux. However, no significant distortion in the light curve shape is observed. The uncertainty due to the index evolution model has been discussed earlier, and different models can change the absolute flux and distort the shape of the light curve. Taking the case of a constant slope as the extreme target, the maximum relative uncertainty is estimated to be around 14\%. Simply taking the root of the sum square of the above uncertainties (as they are correlated), the overall systematic uncertainty is estimated to be 32\% for the flux coefficient $A$ in the fitted light curve parameters. Although the uncertainty may have some influence on other parameters besides $A$, it is expected to be marginal, as the distortion caused by either EBL model or index evolution changes slowly as a function of time, as can be seen from the parameter values under different EBL models in Table~\ref{table:LC-parameter}.

\subsection{Interpretation of the light curve of TeV emission }
The bulk of the prompt  emission of GRB 221009A~\cite{GCN-HEBS,GCN-KW} lasts from $\sim 225$ to $\sim 236$~s, implying that the thickness of the main GRB ejecta is $l\sim c \Delta T$, where $\Delta T\sim 11\, {\rm s}$. The deceleration time $t_{\rm dec}$, represented by the afterglow peak time, is $\sim 18$~s. As $\Delta T\lesssim t_{\rm dec}$, the thin-shell approximation of the ejecta is applicable for the external shock emission~\cite{Sari&Piran1999}. We assume a density profile of the ambient medium to be $n\propto R^{-k}$. $k=0$ corresponds to the case of a homogeneous medium, while $k=2$ corresponds to the case of a stellar wind.

\noindent {\bf -- The rise before the peak}\\
During the coasting phase (before the shock deceleration), the bulk Lorentz factor of the afterglow shock is roughly a constant.   {The shock radius evolves as $R\propto  \Gamma^2 c t \propto  t$, where $\Gamma$ is the Lorentz factor of the shock matter and $t$ is the observer time}. The magnetic field in the shocked ambient medium is $B=\sqrt{(32\pi m_p \epsilon_B n)}\Gamma c\propto R^{-k/2}\propto t^{-k/2}$. The  distribution of the injected electrons by afterglow shock is  assumed to be a power-law $dN_e/d \gamma_e\propto \gamma_e^{-p}$  starting from a minimum random Lorentz factor given by $\gamma_m=\epsilon_e (m_p/m_e)[(p-2)/(p-1)]\Gamma \propto t^0$. The relativistic electrons cool radiatively through synchrotron emission and inverse-Compton (IC) scatterings of the synchroton photons and therefore a cooling break in the electron energy distribution is formed at $\gamma_c=(6\pi m_e c)/[(1+Y_{\rm c})\sigma_{\rm T} \Gamma B^2 t]\propto t^{k-1}$, where $Y_{\rm c}$ is the Compton parameter for electrons with energy $\gamma_c$ and $(1+Y_{\rm c})$  evolves slowly with time. The peak flux of the synchrotron emission is $F_m^{\rm syn}\propto N_e \Gamma B \propto t^{3-3k/2}$, where $N_e\propto R^{3-k}$ is the total number  of swept-up electrons in the postshock medium. The synchrotron emission spectrum has two breaks at the frequencies $\nu_m$ and $\nu_c$, which corresponds to emission frequency of electrons with energy $\gamma_m$ and $\gamma_c$, respectively. The two break frequencies evolve with time as $\nu_m\propto \gamma_m^2 B \propto t^{-k/2}$ and $\nu_c\propto \gamma_c^2 B \propto t^{3k/2-2}$ during the coasting phase.

As an rough approximation, the afterglow SSC spectrum can be described by broken power-laws with two break frequencies at $\nu_{\rm m}^{\rm IC}$ and $\nu_{\rm c}^{\rm IC}$ and a peak flux at $F_{\rm m}^{\rm IC}$ if the IC scattering is in the  Thomson regime, generally resembling the spectrum of the synchrotron emission~\cite{Sari&Esin2001}. The two break frequencies evolves with time as $\nu_m^{\rm IC}=2\gamma_m^2\nu_m\propto t^{-k/2}$ and $\nu_c^{\rm IC}=2\gamma_c^2\nu_c\propto t^{(7k-8)/2}$. The peak flux of the SSC emission is $F_m^{\rm IC}=\tau F_m^{\rm syn}\propto t^{4-5/2k}$, where $\tau$ is the optical depth of the inverse-Compton (IC) scattering, which scales as $\tau \propto N_e/R^2\propto t^{1-k}$. Then the flux is given by
\begin{equation}
F_{\nu}= \left\{
\begin{array}{lll}
F_m^{\rm IC} \left(\frac{\nu}{\nu_m^{\rm IC}}\right)^{-\frac{p-1}{2}}\propto t^{\frac{16-(9+p)k}{4}} \nu^{-\frac{p-1}{2}}, \,\,\,\, \nu_m^{\rm IC}<\nu<\nu_c^{\rm IC}  \\
F_m^{\rm IC} (\frac{\nu}{\nu_c^{\rm IC}})^{-\frac{1}{2}}\propto t^{\frac{8-3k}{4}} \nu^{-1/2},  \,\,\,\, \nu_c^{\rm IC}<\nu<\nu_m^{\rm IC} \\
F_m^{\rm IC} ({\nu_m^{\rm IC}})^{\frac{p-1}{2}}({\nu_c^{\rm IC}})^{\frac{1}{2}} \nu^{-\frac{p}{2}} \propto t^{\frac{8-(2+p)k}{4}} \nu^{-\frac{p}{2}}. \,\,\,\, \nu>\max(\nu_m^{\rm IC}, \nu_c^{\rm IC})
\end{array}
\right.
\end{equation}

Thus, for a wind medium with $k=2$, the fastest rise of the light curve is $t^{1/2}$ in the spectral regime of $F_\nu \propto \nu^{-1/2}$. In other spectral regimes, the flux is either flat or decays with time from the start. The TeV spectrum of GRB 221009A does not agree with $F_\nu \propto \nu^{-1/2}$, so we expect  a flat or declining light curve in the wind medium scenario. On the other hand, for a homogeneous medium with $k=0$, the flux rises with time as $t^{2}$ in the spectral regime of $F_{\nu} \propto \nu^{-p/2}$ (i.e., $\nu_m^{\rm IC}<\nu_c^{\rm IC}<\nu$). The observed slope of $\alpha_1 = 1.82_{-0.18}^{+0.21}$ for GRB 221009A, together with a soft spectrum $\gamma_{\rm VHE}>2$ (implying $\nu>\max \left\{\nu_m^{\rm IC}, \nu_c^{\rm IC}\right\}$), is consistent with the this scenario, so we infer $k=0$.

At the very early phase, the flux could rise  as $t^{4}$ if the spectrum is in the  regime $F_{\nu} \propto \nu^{-(p-1)/2}$ (i.e., $\nu_m^{\rm IC}<\nu<\nu_c^{\rm IC}$), which is possible as $\nu_c^{\rm IC} \propto t^{-4}$ (for $k=0$). In addition, this phase overlaps with the strongest pulse of the prompt main burst emission~\cite{GCN-HEBS,GCN-KW}. The inner ejecta could catch up with the external shock  driven by the outer ejecta and add energy to the external shock. This  energization effect could  also increase the bulk Lorentz factor of the external shock and hence increases the TeV flux dramatically (note that $F_m^{\rm IC}$ strongly depends on $\Gamma$).  This could explain the very rapid rise of the TeV emission from $T_0+230 \,{\rm s}$ to $T_0+231\,{\rm s}$ in GRB 221009A. 

\noindent {\bf -- The decay after the peak}\\
After the peak, the shock is decelerated by the swept-up ambient mass.
For a homogeneous medium, the bulk Lorentz factor decreases with time as $\Gamma\propto t^{-3/8}$. Then one has $\nu_m^{\rm IC} \propto t^{-9/4}$, $\nu_c^{\rm IC} \propto t^{-1/4}$, and $F_m^{\rm IC}\propto \tau F_m^{\rm syn}$. {As $\tau=\frac{1}{3} n\sigma_{\rm T}R\propto t^{1/4}$}, we get $F_m^{\rm IC}\propto t^{1/4}$. Thus, for a homogeneous medium~\cite{Wang2019},
\begin{equation}
\label{decay-slope}
F_{\nu}= \left\{
\begin{array}{lll}
F_m^{\rm IC} \left(\frac{\nu}{\nu_m^{\rm IC}}\right)^{-\frac{p-1}{2}}\propto t^{\frac{11-9p}{8}}, \,\,\,\, \nu_m^{\rm IC}<\nu<\nu_c^{\rm IC}  \\
F_m^{\rm IC} (\frac{\nu}{\nu_c^{\rm IC}})^{-\frac{1}{2}}\propto t^{\frac{1}{8}},  \,\,\,\, \nu_c^{\rm IC}<\nu<\nu_m^{\rm IC} \\
F_m^{\rm IC} ({\nu_m^{\rm IC}})^{\frac{p-1}{2}}({\nu_c^{\rm IC}})^{\frac{1}{2}} \nu^{-\frac{p}{2}}\propto t^{\frac{10-9p}{8}}, \,\,\,\, \nu>\max(\nu_m^{\rm IC}, \nu_c^{\rm IC})
\end{array}
\right.
\end{equation}

As the Klein-Nishina (KN) effect becomes more important at later times, the spectral regime could enter into the KN regime at later time for  SSC emission. The KN effect does not lead to a sharp cutoff but a consecutive set of power-law segments that become steeper at higher frequencies~\cite{Nakar2009}. A critical Lorentz factor is~\cite{Nakar2009}
\begin{equation}
\hat{\gamma}_c=\frac{\Gamma m_e c^2}{h\nu_c},
\end{equation}
corresponding to the critical electrons that up-scattering photons at energy $h\nu_c$ in the Thompson scattering regime. The KN limit affect the SSC spectrum  at $\nu>2\nu_c{\hat\gamma}_c \max(\gamma_c,\hat{\gamma}_c)$. If $\gamma_c<\hat{\gamma}_c$, the total SSC luminosity is not  suppressed by the KN limit and the SSC peak is
observed at $h\nu^{\rm IC}_{\rm c}=2\gamma_c^2h\nu_c$. 
If $\gamma_c>\hat{\gamma}_c$, the SSC peak is observed at  $h\nu^{\rm IC}_{\rm c,KN}=2\xi_{\rm KN}\gamma_c \hat{\gamma}_c h\nu_c=2\xi_{\rm KN}\Gamma \gamma_c m_e c^2$, where  $\xi_{\rm KN}=0.1$ is a correction factor taking into account the onset of KN corrections at energies less than $m_e c^2$, which is needed in comparison with the numerically-reproduced SSC spectra that uses the exact cross-section~\cite{Yamasaki2022}.
In the latter case, the flux immediately above $\nu^{\rm IC}_{\rm c,KN}$ is given by
\begin{equation}
F_{\nu}=F_m^{\rm IC} ({\nu_m^{\rm IC}})^{\frac{p-1}{2}}({\nu^{\rm IC}_{\rm c,KN})^{\frac{p-1}{2}} \nu^{-(p-1)} \propto t^{({\frac{3}{2}-\frac{5p}{4}})} \nu^{-(p-1)}}. 
\end{equation}
This spectral index  is almost not distinguishable from the $-p/2$ spectral index expected in case that KN effects are negligible if $p\simeq 2$.
The observed temporal slope of the TeV flux after the peak is consistent with $t^{\frac{10-9p}{8}}\sim t^{-1.1}$ for $p\sim2.1$, if the spectrum is within  the spectral regime of $F_\nu\sim \nu^{-\frac{p}{2}}$. It is also consistent with $\propto t^{({\frac{3}{2}-\frac{5p}{4}})}$ for $p\sim2.1$ in the case that the KN effect is strong, where the spectrum is  $F_\nu\sim \nu^{-(p-1)}$.

\noindent {\bf -- The light curve steepening }\\
The light curve steepening at $t_{\rm b,2}\simeq 670 \,{\rm s}$ after $T^*$ cannot be due to the KN effect, because the spectrum after the break does not  soften. In addition, the steepening due to the KN effect would have a break time dependent on the frequency, which is inconsistent with the observations.  The  cross of a spectral break to the observed frequency sometimes leads to a temporal break. However, for a homogeneous medium,  since $\nu_c^{\rm IC}$ decrease with time, we do not expect temporal break due to the spectral transition according to Eq.~\ref{decay-slope}.

The steepening could be due to the jet effect, in which the jet edge is seen by the observer. Due to relativistic beaming, only a small part of the jet with angular size of $ \Gamma^{-1}$ is  visible to the observer initially, where $\Gamma$ is the Lorentz factor of the bulk motion of the emitting material. Thus the
observer is unable to distinguish a sphere from a jet as long as $\Gamma>1/\theta_0$, where $\theta_0$ is the initial half opening angle of the jet.  However, as the shock continues its radial expansion, $\Gamma$ will decrease, and when $\Gamma<1/\theta_0$, the observer will see the flux reduced by the ratio of the solid angle of the emitting surface $\theta_0^2$ to that expected for a spherical expansion ($1/\Gamma^{2}$), leading to a steepening by $\Gamma^2\theta_0^2\propto t^{-3/4}$. 

The jet could also expand sideways when $\Gamma$ decreases below $1/\theta_0$~\cite{Rhoads1999,Sari1999}. Consequently, in this regime, 
$\Gamma$ decreases exponentially with radius $R$ instead of as a power law. We can therefore take
$R$ to be essentially constant during this spreading phase.  Since $t\sim R/2\Gamma^2 c$, we find $\Gamma \propto t^{-1/2}$ during the spreading phase. The magnetic field in the shocked ambient medium is $B\propto n^{1/2}\Gamma\propto t^{-1/2}$.
The minimum Lorentz factor of the relativistic electrons is $\gamma_m\propto \Gamma\propto t^{-1/2}$. Then we have $\nu_m\propto \Gamma \gamma_m^2 B\propto t^{-2}$ and $\nu_m^{\rm IC}\propto \gamma_m^2 \nu_m\propto t^{-3}$. The cooling break of the relativistic electrons is $\gamma_c\propto 1/(\Gamma B^2 t)\propto t^{1/2}$. Then the cooling frequencies in the synchrotron and SSC spectra are $\nu_c\propto \Gamma \gamma_c^2 B\propto t^0$ and $\nu_c^{\rm IC}=2\gamma_c^2\nu_c\propto t$. The peak flux of the SSC spectrum is $F_m^{\rm IC}\propto \tau F_m^{\rm syn}\propto t^{-1}$. Then we have
\begin{equation}
\label{jet-slope}
F_{\nu}= \left\{
\begin{array}{lll}
F_m^{\rm IC} \left(\frac{\nu}{\nu_m^{\rm IC}}\right)^{-\frac{p-1}{2}}\propto t^{-\frac{3p-1}{2}}, \,\,\,\, \nu_m^{\rm IC}<\nu<\nu_c^{\rm IC}  \\
F_m^{\rm IC} (\frac{\nu}{\nu_c^{\rm IC}})^{-\frac{1}{2}}\propto t^{-\frac{1}{2}},  \,\,\,\, \nu_c^{\rm IC}<\nu<\nu_m^{\rm IC} \\
F_m^{\rm IC} ({\nu_m^{\rm IC}})^{\frac{p-1}{2}}({\nu_c^{\rm IC}})^{\frac{1}{2}} \nu^{-\frac{p}{2}}\propto t^{-\frac{3p-2}{2}}, \,\,\,\, \nu>\max(\nu_m^{\rm IC}, \nu_c^{\rm IC})
\end{array}
\right.
\end{equation}

If $\gamma_c>\hat{\gamma}_c$, the SSC peak is observed at  $h\nu^{\rm IC}_{\rm c,KN}=2\xi_{\rm KN}\gamma_c \hat{\gamma}_c h\nu_c=2\xi_{\rm KN}\Gamma \gamma_c m_e c^2$, which evolves as $t^0$ for the sideways expansion phase.  The flux above $\nu^{\rm IC}_{\rm c,KN}$ is given by
\begin{equation}
F_{\nu}=F_m^{\rm IC} ({\nu_m^{\rm IC}})^{\frac{p-1}{2}}({\nu^{\rm IC}_{\rm c,KN})^{\frac{p-1}{2}} \nu^{-(p-1)} \propto t^{-(3p-1)/2} \nu^{-(p-1)}}. 
\end{equation}

\subsection{Multi-wavelength light curve analysis and modeling}

\begin{figure}
\centering
\includegraphics[width=0.8\linewidth]{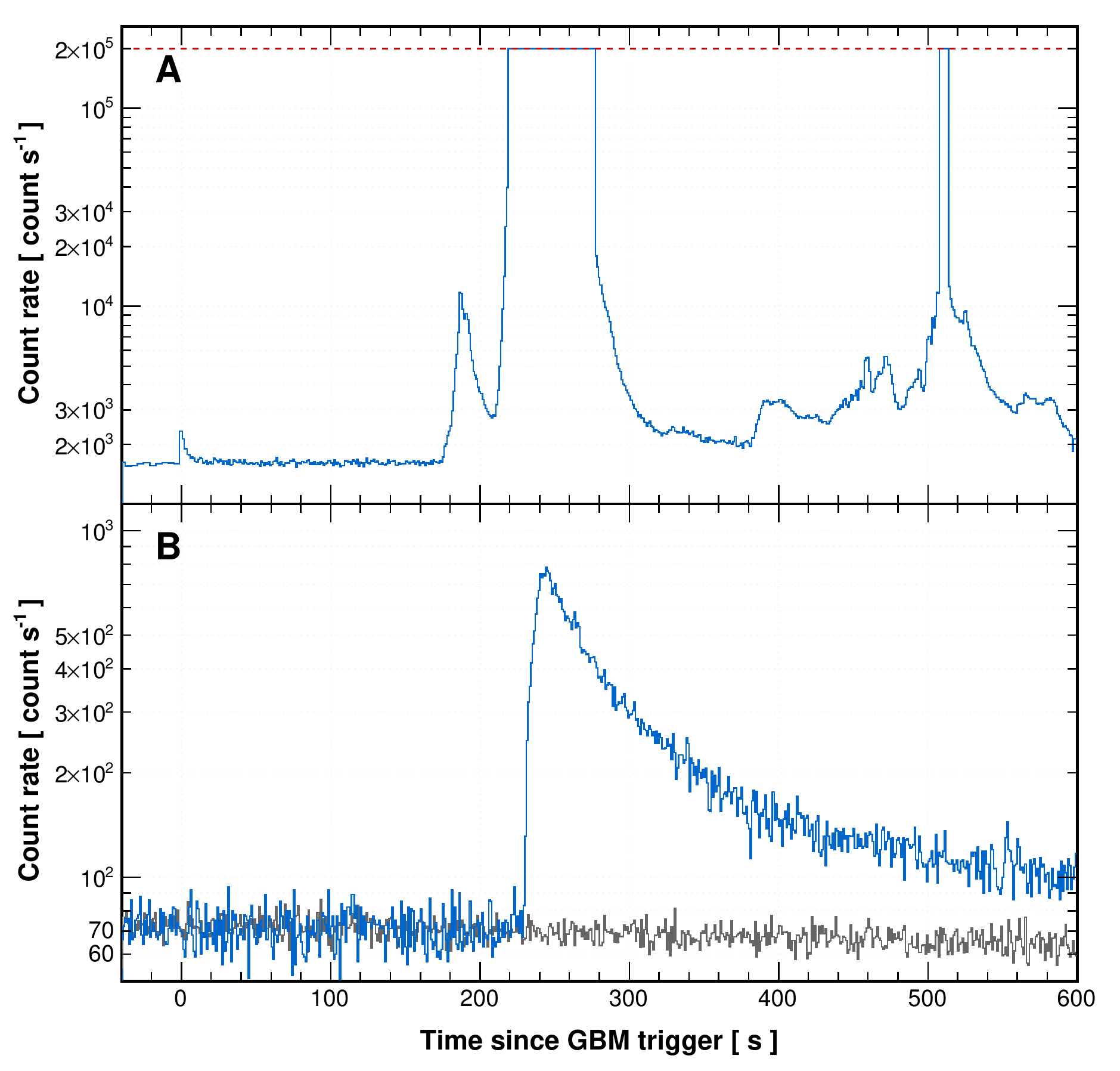}
\caption{{\bf Comparison between the keV--MeV light curve measured by Fermi/GBM and the TeV light curve measured by LHAASO-WCDA.} (A) the count-rate light curve of 200~keV--40~MeV emission measured by Fermi/GBM (BGO detector). The red horizontal dashed line indicates the level at which the detector became saturated during two periods: 219--277 and 508--514 seconds after the GBM trigger. (B) the count-rate light curve (in blue) of GRB 221009A with $N_{\rm hit} \geq 30$ (energy range 0.2--7~TeV) detected by LHAASO-WCDA in the first $\sim 600$ seconds, while the black curve shows the background rates.}
\label{fig:rate_gbm_wcda}
\end{figure}

\noindent{\bf -- {Fermi}/GBM light curve}\\
The GBM data of this GRB were downloaded from the Fermi/GBM public data archive~\cite{GbmDataTools}.  
We first used the time-tagged event (TTE) data to estimate the time interval of the pile-up effect, and found that the count rates were affected by pulse pile-up at two time intervals $\sim \rm 219-277~s$ and $\sim \rm 508-514~s$ after the GBM trigger. 
Based on the GBM Data Tools~\cite{GbmDataTools}, the light curve in the energy band from 200 keV to 40 MeV, shown in Fig.~\ref{fig:rate_gbm_wcda}, was obtained using the Continuous Spectroscopy data for the brightest bismuth germanate (BGO) scintillator. 

\noindent {\bf -- {Swift}/XRT light curve}\\
The \textit{Swift}/XRT data were obtained from the Swift online repository~\cite{Evans2007}. The temporally resolved energy-flux light curve shown in Fig.~\ref{fig:modeling} was derived using the online Burst Analyser tool~\cite{Evans2007}. We performed the spectral analysis of the time interval 170--2000~s after the \textit{Swift}/BAT trigger and estimated  a photon index of $\Gamma_{\rm X}=-1.65\pm0.02$.  We  use a conversion factor from counts to flux equal to $6.24 \times 10^{-11} \, {\rm erg \, cm^{-2} \, count^{-1}}$. This conversion factor was applied to the counts light curve to derive the energy flux light curve in the time interval 170--2000~s.

\noindent {\bf -- Modeling}\\
We  consider that a jet with isotropically-equivalent kinetic energy  $E_k$ and a half opening angle  $\theta_0$ drives a forward shock expanding into a homogeneous medium with number density  $n$. The modeling of the multi-wavelength emission is based on the numerical code that has been applied to previous GRB afterglows~\cite{Liu2013,Wang2019}.  This code uses shock dynamics  taken from~\cite{Huang1999}.  The radius of forward shock where the emission is produced is calculated by $dR/dt=\beta c/(1-\beta)$, where $R$ is the shock radius and $\beta$ is the shock velocity in unit of the speed of light. We consider only the jet edge effect on the light curve without taking into account sideways expansion. 
We assume that electrons swept up by the shock are accelerated into a power law distribution described by spectral index $p$: $dN/d\gamma_{\rm e}\propto \gamma^{-p}_{\rm e}$ , where $\gamma_{\rm e}$ is the electron Lorentz factor. A full Klein-Nishina cross section for the inverse Compton scattering has been used and the $\gamma\gamma$ absorption has been taken into account.  The KN effect may also affect the electron distribution, so we have calculated the electron distribution self-consistently.

The absorption optical depth  due to internal $\gamma\gamma$ interactions is given by 
\begin{equation}
\label{ggabsorb}
    \tau_{\gamma\gamma,\rm int}(\varepsilon_{\gamma},\varepsilon_{\rm t})=\int_{\varepsilon_{\rm th}}\sigma_{\gamma\gamma}(\varepsilon_{\gamma},\varepsilon_{\rm t})\frac{R}{\Gamma}\frac{dn}{d\varepsilon_{\rm t}}d\varepsilon_{\rm t}d\Omega,
\end{equation}
where $\sigma_{\gamma\gamma}$ is the cross section, {$\varepsilon_{\rm th}=2m^2_{\rm e}c^4/\varepsilon_{\gamma}(1-{\rm cos}\theta)$ is the threshold energy} and $dn/d\varepsilon_{\rm t}$ is the differential number density of synchrotron target photons. The cross section is given by
\begin{equation}
    \sigma_{\gamma\gamma}(\varepsilon_{\gamma},\varepsilon_{\rm t})=\frac{3}{{16}}{\sigma _{\rm T}}\left( {1 - {\beta^2_{\gamma\gamma}}} \right)\left[ {2\beta_{\gamma\gamma} \left( {{\beta^2_{\gamma\gamma}} - 2} \right) + \left( {3 - {\beta^4_{\gamma\gamma}}} \right)\ln \left( {\frac{{1 + \beta_{\gamma\gamma} }}{{1 - \beta_{\gamma\gamma} }}} \right)} \right],
\end{equation}
where {$\beta_{\gamma\gamma}=\sqrt{1-2m^2_{\rm e}c^4/\varepsilon_{\gamma}\varepsilon_{\rm t}(1-{\rm cos}\theta)}$}.

We can use an approximate analytic approach to obtain an estimate of the shock parameter values. 
Characteristic frequency $\nu_m$ and $\nu_c$ represent the characteristic synchrotron frequency of electrons with minimum Lorentz factor $\gamma_{\rm m}$  and the cooling Lorentz factor $\gamma_{\rm c}$, respectively. {The X-ray spectral index measured by XRT is $\beta_{\rm X}=-1.65^{+0.02}_{-0.02}$. This can be explained by $\beta_{\rm X}=-(p-1)/2 $with $p=2.3\pm0.04$ in the regime of $\nu_m<\nu_{\rm X}<\nu_c$. Thereafter, we take a slow-cooling spectrum for the afterglow of GRB 221009A.}

The emission at $\sim 100$ MeV measured by Fermi/LAT is thought to be produced by  synchrotron emission above $\nu_{\rm c}$. 
The flux  is given by~\cite{Kumar2009}
\begin{equation}
    F({\rm 100MeV})=3\times10^{-6}E^{\frac{p+2}{4}}_{k,55}\epsilon^{p-1}_{\rm e,-1}\epsilon^{\frac{p-2}{4}}_{\rm B,-4}t^{-\frac{3p-2}{4}}_{2} {\rm erg\ cm^{-2}\ s^{-1}}.
\end{equation}
By comparing it with the measured flux of $F (100\,{\rm MeV})\sim 7.3\times10^{-7}\,{\rm erg\ cm^{-2}\ s^{-1}}$ from $T^*$ + 68~s to $T^*$ + 174~s~\cite{Liu2022}, we obtain $\epsilon_{\rm e}\sim 0.036E_{\rm k,55}^{-0.8}$ for $p=2.3$, considering that the flux is insensitive to $\epsilon_{\rm B}$.  The X-ray flux observed by XRT in post jet-break phase at $T^*+3000{\rm s}$ can set upper limits on the values of $\epsilon_{\rm B}$ and $n$. Assuming the spectra regime is $\nu_{\rm m}<\nu_{\rm X}(1 {\rm keV}) <\nu_{\rm c}$, we find that the flux at 1 keV is $F(1{\rm keV})=2.3\times10^{-8} E^{\frac{p+4}{4}}_{k,55}\epsilon^{p-1}_{\rm e,-1}n^{1/4}_{0}\epsilon^{\frac{p+1}{4}}_{\rm B,-4}t^{\frac{-3p}{4}}_{3.5} {\rm ergcm^{-2}s^{-1}}$ after considering the light curve steepening (by a factor of $3/4$ in the decay slope) at the jet break time $670{\rm s}$. This flux should not be greater than the XRT flux, i.e.,   $F(1 {\rm keV})\lesssim4\times10^{-8}{\rm ergcm^{-2}s^{-1}}$, leading to $n^{1/4}_{0}\epsilon^{0.83}_{\rm B,-4}\lesssim6.7E_{k,55}^{-1.55}$ (for $p=2.3$).


The spectra of LHAASO observations can be used to constrain  the  SSC peak energy $E_{\rm peak}$ and the peak flux $F^{\rm IC}_{\rm m}$. The peak energy should be $E_{\rm peak}=\min \left\{E_{\rm c}^{\rm IC},E_{\rm c, KN}^{\rm IC}\right\}$, where  $E^{\rm IC}_{\rm c, KN}=0.8E^{-1/4}_{k,55}n^{-3/4}_{0}\epsilon^{-1}_{\rm B,-4}t^{-1/4}_{2}$TeV in the KN regime   and  $E^{\rm IC}_{\rm c}=8.8E^{-5/4}_{k,55}n^{-9/4}_{0}\epsilon^{-7/2}_{\rm B,-4}t^{-1/4}_{2}$TeV in the Thomson regime. 
The LHAASO spectra require $E_{\rm peak}\lesssim300 {\rm GeV}$ at $T^*+100{\rm s}$, leading to $n^{9/4}_{0}\epsilon^{7/2}_{\rm B,-4}\gtrsim29E^{-5/4}_{k,55}$ in the Thomson regime and $n^{3/4}_{0}\epsilon_{\rm B,-4}\gtrsim2.6E^{-1/4}_{k,55}$ in the KN regime.
The peak flux density,  given by $F^{\rm IC}_{\rm m}=1.4E^{5/4}_{k,55}n^{5/4}_{0}\epsilon^{1/2}_{\rm B,-4}t^{1/4}_{2}  \mu$Jy, should be greater than the  observed flux $F_{\nu}(300\,{\rm GeV})\sim 0.013\mu$Jy, leading to $n^{5/4}_{0}\epsilon^{1/2}_{\rm B,-4}\gtrsim0.01E^{-5/4}_{k,55}$. 

There is still degeneracy among the parameter values,  so we choose the following set of  parameter values  satisfying all the above constraints:  $E_{k}= 1.5\times10^{55}\,{\rm erg}$, $\epsilon_e= 0.025$,  $n= 0.4\, {\rm cm^{-3}}$, $\epsilon_B= 6\times 10^{-4}$,  $p= 2.2$,  $\Gamma_0= 560$   and  $\theta_0= 0.8^\circ$. 
For these parameter values, we find the internal optical depth obtained  is $\tau_{\gamma\gamma}\leq1$ at $\rm 5 TeV$.

We summarize the spectral regimes of the TeV emission for the above set of parameter values as follows: The electron distribution at all phases is  slow-cooling  with $\gamma_m<\gamma_c$.
During the coasting phase, the KN effect is not important as $E_{\rm c, KN}^{\rm IC}$ remains above the observed frequency until $t\sim T^{*}$ + 20 s (i.e., $E_{\rm c, KN}^{\rm IC}>h\nu_{\rm obs}>E_c^{\rm IC}>E_m^{\rm IC}$). The SSC flux rises as $F_{\rm \nu}\sim t^{2}$ with a spectrum $F_\nu\propto \nu^{-p/2}$. Afterwards, as $E^{\rm IC}_{\rm c,KN}$ decreases and falls below 300~GeV, the observed frequency enters the KN regime, leading to  $F_{\rm \nu} \propto t^{({\frac{3}{2}-\frac{5p}{4}})}$ with a spectrum $F_\nu\propto \nu^{-(p-1)}$ ($E_{\rm c}^{\rm IC}>h\nu_{\rm obs}>E_{\rm c,KN}^{\rm IC}>E_m^{\rm IC}$). After the jet break, as we only consider the jet edge effect, the temporal slope steepens by a factor of $\frac{3}{4}$ while the spectrum remains unchanged.

\subsection{The SSC emission in the prompt emission phase}
We study the synchrotron self-Compton (SSC) emission from internal shocks in the scenario that  the prompt MeV emission of GRBs is produced by the synchrotron emission. The Lorentz factor of the flow, $\Gamma_0$, is assumed to vary on a
timescale $t_v$  with a typical amplitude of $\Delta \Gamma\sim \Gamma_0$.  During the rising phase of the main burst emission  ($T_0+210\,{\rm s}$ to $T_0+219\,{\rm s}$), when the detector is not saturated,  the shortest variability time  is found to be $t_v=0.082\, {\rm s}$~\cite{Liu2022}.  The
collisions between the shells typically occur at a radius   $R_{\rm in}\sim 2\Gamma_0^2 c  t_v=10^{15}\,{\rm cm}(\Gamma_0/440)^2 ( t_v/0.082\, {\rm s})$. If there is variability on
a shorter/longer timescale, this would lead a smaller/larger radius for the internal dissipation, which scales linearly with  $t_v$.

TeV gamma-rays will suffer from pair production absorption with low-energy photons in the prompt emission. For a photon energy with $\varepsilon_\gamma$,  the target photons for the annihilation is $\varepsilon_t=2\Gamma^2 (m_e c^2)^2/\varepsilon_\gamma=100\,{\rm keV} (\Gamma_0/440)^2 (\varepsilon_\gamma/{\rm 1\, TeV})^{-1}$. The prompt emission has a Band function with index of $\beta_1$ and $\beta_2$ below and above the break energy~\cite{Band}. Then the luminosity of photons at $\varepsilon_t$ is $L(\varepsilon_t)=L_{\gamma} (\varepsilon_t/h\nu_m)^{2+\beta_1}$. Then the number density of photons  at $\varepsilon_t$ in the co-moving frame is  $n'_t=L(\varepsilon_t)/(4\pi R_{\rm in}^2 \Gamma_0 c \varepsilon_t)$. The $\gamma\gamma$ absorption optical depth is
\begin{equation}
\tau_{\gamma\gamma}\sim \sigma_{\gamma\gamma}n'_t \frac{R_{\rm in}}{\Gamma_0}\sim 190\left(\frac{R_{\rm in}}{10^{15}\,{\rm cm}}\right)^{-1} \left(\frac{\Gamma_0}{440}\right)^{-2} \left(\frac{\varepsilon_t}{h\nu_m}\right)^{\beta_1+1},
\end{equation}
where we have used $L_\gamma=L(h\nu=1\,{\rm MeV})=3\times10^{53}\,{\rm erg\, s^{-1}}$ and $\sigma_{\gamma\gamma}\simeq 0.15\sigma_{\rm T}$ (the value of two-photon pair production cross-section near its peak) in the calculation. For $\beta_1\sim -1$ and $\Gamma_0\sim 440$, the optical depth is estimated to be  $\tau_{\gamma\gamma}\sim 190\left({R_{\rm in}}/{10^{15}\,{\rm cm}}\right)^{-1}$, which  results in strong attenuation of TeV photons,  explaining the  low flux ratio between TeV  and  MeV emission.  

\clearpage

\begin{table}
\setlength{\tabcolsep}{4pt} 
\renewcommand{\arraystretch}{1.2} 
\begin{center}   
\caption{{\bf Number of events in each $\bm{N_{\rm hit}}$ bin.} The eight columns marked with numerical ranges are $N_{\rm hit}$ bins. The data types are: {\sl OBS} the number of observed events; {\sl ERR} the statistical uncertainty of the number of observed events; {\sl PL} the number of expected events based on the intrinsic power-law fitting (as shown in Fig.~\ref{fig:SED}A); {\sl SCC} the number of expected events based on the intrinsic SSC modeling (as shown in Fig.~\ref{fig:modeling}); {\sl ME} the median energy (in TeV) obtained from the simulation using the fitted intrinsic power-law function as SED.}
\label{table:spectrum-events}
\scriptsize

\begin{tabular}{ccccccccccc} 
\toprule

Time interval & Type & [30, 40) & [40, 63) & [63, 100) & [100, 160) & [160, 250) & [250, 400) & [400, 500) & [500, 1000) & $\chi^2/{\rm dof}$\\
{\scriptsize (seconds after $T_0$)} & & & & & & & & & &\\
\midrule
\multirow{5}*{231--240} & {\sl OBS} & $206.3$ & $252.2$ & $209.8$ & $110.5$ & $50.0$ & $21.0$ & $3.0$ & $2.0$ & $\backslash$\\
 & {\sl ERR} & $14.7$ & $15.9$ & $14.5$ & $10.5$ & $7.1$ & $4.6$ & $1.7$ & $1.4$ & $\backslash$\\
 & {\sl PL} & $214.4$ & $232.4$ & $215.0$ & $118.0$ & $52.1$ & $17.2$ & $2.5$ & $1.0$ & $3.1/6$\\
 & {\sl SSC} & $180.7$ & $232.4$ & $244.8$ & $153.6$ & $72.2$ & $23.9$ & $3.4$ & $1.3$ & $26.9/8$\\
 & {\sl ME} & $0.234$ & $0.338$ & $0.515$ & $0.919$ & $1.749$ & $2.996$ & $3.963$ & $5.203$ & $\backslash$\\
\midrule
\multirow{5}*{240--248} & {\sl OBS} & $329.2$ & $329.5$ & $338.5$ & $169.8$ & $62.0$ & $33.0$ & $4.0$ & $3.0$ & $\backslash$\\
 & {\sl ERR} & $18.4$ & $18.3$ & $18.4$ & $13.0$ & $7.9$ & $5.7$ & $2.0$ & $1.7$ & $\backslash$\\
 & {\sl PL} & $323.6$ & $347.1$ & $317.7$ & $171.6$ & $74.6$ & $24.2$ & $3.5$ & $1.4$ & $6.5/6$\\
 & {\sl SSC} & $258.0$ & $321.2$ & $329.8$ & $200.8$ & $92.5$ & $30.2$ & $4.3$ & $1.7$ & $25.5/8$\\
 & {\sl ME} & $0.230$ & $0.333$ & $0.507$ & $0.907$ & $1.734$ & $2.983$ & $3.952$ & $5.194$ & $\backslash$\\
\midrule
\multirow{5}*{248--326} & {\sl OBS} & $1345.3$ & $1743.3$ & $1624.1$ & $936.5$ & $397.8$ & $138.5$ & $28.0$ & $6.0$ & $\backslash$\\
 & {\sl ERR} & $37.9$ & $42.1$ & $40.4$ & $30.7$ & $20.0$ & $11.8$ & $5.3$ & $2.4$ & $\backslash$\\
 & {\sl PL} & $1465.4$ & $1652.6$ & $1578.3$ & $928.9$ & $422.1$ & $143.4$ & $20.9$ & $8.2$ & $8.7/6$\\
 & {\sl SSC} & $1330.1$ & $1644.3$ & $1665.9$ & $1020.5$ & $454.8$ & $144.5$ & $19.4$ & $7.2$ & $13.0/8$\\
 & {\sl ME} & $0.239$ & $0.359$ & $0.554$ & $0.955$ & $1.826$ & $2.929$ & $4.213$ & $5.374$ & $\backslash$\\
\midrule
\multirow{5}*{326--900} & {\sl OBS} & $1248.8$ & $1572.1$ & $1582.4$ & $1094.3$ & $497.8$ & $189.0$ & $21.0$ & $4.6$ & $\backslash$\\
 & {\sl ERR} & $42.7$ & $41.8$ & $40.8$ & $33.3$ & $22.4$ & $13.8$ & $4.6$ & $2.3$ & $\backslash$\\
 & {\sl PL} & $1278.2$ & $1584.8$ & $1556.3$ & $1066.2$ & $506.6$ & $188.6$ & $22.7$ & $8.5$ & $3.4/6$\\
 & {\sl SSC} & $1179.8$ & $1497.5$ & $1442.3$ & $915.7$ & $380.3$ & $119.6$ & $11.5$ & $4.0$ & $70.9/8$\\
 & {\sl ME} & $0.286$ & $0.399$ & $0.631$ & $1.140$ & $1.969$ & $2.949$ & $4.565$ & $5.616$ & $\backslash$\\
\midrule
\multirow{5}*{900--2000} & {\sl OBS} & $263.3$ & $284.5$ & $265.0$ & $177.9$ & $77.8$ & $34.2$ & $2.0$ & $\backslash$ & $\backslash$\\
 & {\sl ERR} & $33.3$ & $23.2$ & $19.1$ & $14.4$ & $9.3$ & $6.0$ & $1.4$ & $\backslash$ & $\backslash$\\
 & {\sl PL} & $249.2$ & $288.2$ & $271.4$ & $173.7$ & $80.7$ & $28.5$ & $3.4$ & $\backslash$ & $2.2/5$\\
 & {\sl SSC} & $235.1$ & $278.8$ & $259.0$ & $154.4$ & $63.6$ & $19.0$ & $1.9$ & $\backslash$ & $11.7/7$\\
 & {\sl ME} & $0.301$ & $0.411$ & $0.656$ & $1.260$ & $2.059$ & $3.171$ & $4.596$ & $\backslash$ & $\backslash$\\

\bottomrule
\end{tabular}
\end{center}
\end{table}

\begin{table}
\begin{center}   
\caption{{\bf Spectrum parameters for the selected time intervals.} The observed spectra are fitted using the function ${\rm d}N/{\rm d}E = A (E/{\rm TeV})^{-\gamma}\,e^{-E/E_{\rm cut}}$, where $E_{\rm cut} = 3.14 \pm 0.41\,{\rm TeV}$ is obtained by fitting the events in the full time range (231--2000~s after $T_0$). The intrinsic spectrum is fitted using ${\rm d}N/{\rm d}E = A (E/{\rm TeV})^{-\gamma}$. Intrinsic spectrum parameters under two other EBL models, corresponding to the upper and the lower error boundary in the model~\cite{Saldana-Lopez2021}, are also listed, marked with high and low, respectively, for comparison with the parameters obtained using the standard model.}
\label{table:spectrum-parameter}
\small
\begin{tabular}{ccccc} 
\toprule
Time interval   & $A$ & $\gamma$ & $E_{\rm cut}$ & $\chi^2/{\rm dof}$ \\
(seconds after $T_0$) & ($10^{-8}\,{\rm TeV}^{-1}\,{\rm cm}^{-2}\,{\rm s}^{-1}$) & & (TeV) &  \\

\midrule
\multicolumn{5}{c}{Observed spectrum}\\
\midrule
231--240 & $42.9 \pm 2.7$ & $2.983 \pm 0.061$ & $3.14$ (fixed) & $4.6/6$ \\
240--248 & $70.1 \pm 3.8$ & $3.006 \pm 0.052$ & $3.14$ (fixed) & $8.0/6$ \\
248--326 & $39.9 \pm 1.0$ & $2.911 \pm 0.028$ & $3.14$ (fixed) & $14.8/6$ \\
326--900 & $7.35 \pm 0.16$ & $2.788 \pm 0.026$ & $3.14$ (fixed) & $8.9/6$ \\
900--2000 & $0.959 \pm 0.043$ & $2.880 \pm 0.067$ & $3.14$ (fixed) & $2.9/5$ \\

\midrule
\multicolumn{5}{c}{Intrinsic spectrum, standard EBL}\\
\midrule
231--240 & $127.3 \pm 7.9$ & $2.429 \pm 0.062$ & $\backslash$ & $3.1/6$ \\
240--248 & $208 \pm 11$ & $2.455 \pm 0.054$ & $\backslash$ & $6.5/6$ \\
248--326 & $117.8 \pm 3.0$ & $2.359 \pm 0.028$ & $\backslash$ & $8.7/6$ \\
326--900 & $21.77 \pm 0.47$ & $2.231 \pm 0.026$ & $\backslash$ & $3.4/6$ \\
900--2000 & $2.84 \pm 0.13$ & $2.324 \pm 0.065$ & $\backslash$ & $2.2/5$ \\

\midrule
\multicolumn{5}{c}{Intrinsic spectrum, high EBL}\\
\midrule
231--240 & $164 \pm 11$ & $2.295 \pm 0.066$ & $\backslash$ & $3.5/6$ \\
240--248 & $267 \pm 15$ & $2.322 \pm 0.055$ & $\backslash$ & $7.1/6$ \\
248--326 & $152.5 \pm 4.0$ & $2.217 \pm 0.029$ & $\backslash$ & $7.6/6$ \\
326--900 & $28.23 \pm 0.61$ & $2.082 \pm 0.027$ & $\backslash$ & $1.3/6$ \\
900--2000 & $3.68 \pm 0.16$ & $2.172 \pm 0.067$ & $\backslash$ & $2.1/5$ \\

\midrule
\multicolumn{5}{c}{Intrinsic spectrum, low EBL}\\
\midrule
231--240 & $98.7 \pm 6.0$ & $2.564 \pm 0.061$ & $\backslash$ & $2.6/6$ \\
240--248 & $161.2 \pm 8.5$ & $2.589 \pm 0.052$ & $\backslash$ & $5.9/6$ \\
248--326 & $90.7 \pm 2.3$ & $2.505 \pm 0.027$ & $\backslash$ & $11.8/6$ \\
326--900 & $16.71 \pm 0.35$ & $2.385 \pm 0.025$ & $\backslash$ & $9.4/6$ \\
900--2000 & $2.188 \pm 0.098$ & $2.478 \pm 0.062$ & $\backslash$ & $2.6/5$ \\

\bottomrule
\end{tabular}
\end{center}
\end{table}

\begin{table}
\begin{center}   
\caption{
{\bf Fitted parameters for light curves at different $\bm{N_{\rm hit}}$ segments}: The row labeled mask is for the case where the flare period is masked. The row labeled low or high is for the case where the low or the high EBL model is used. The reference time $T^* = T_0 + 226\,{\rm s}$ is used for the fitting. The parameter $t_{\rm peak}$ is derived from other parameters, with  $t_{\rm peak} = t_{\rm b,1}\left(-\alpha_1/\alpha_2\right)^{1/\left(\omega_1(\alpha_1-\alpha_2)\right)}$.}
\label{table:LC-parameter} 
\scriptsize
\linespread{1.3}\selectfont
\newcommand{\entrya}[1]{\multicolumn{1}{m{0.12\linewidth}}{\centering #1}}
\newcommand{\entryb}[1]{\multicolumn{1}{m{0.08\linewidth}}{\centering #1}}
\newcommand{\entryc}[1]{\multicolumn{1}{m{0.16\linewidth}}{\centering #1}}
\newcommand{\entryd}[1]{\multicolumn{1}{m{0.16\linewidth}}{\centering #1}}
\newcommand{\entrye}[1]{\multicolumn{1}{m{0.16\linewidth}}{\centering #1}}
\newcommand{\entryf}[1]{\multicolumn{1}{m{0.16\linewidth}}{\centering #1}}
\renewcommand\arraystretch{1.2}
\begin{tabular}{c c c c c c}\toprule
    \entrya{$N_{\rm hit}$\\segment}
  & \entryb{$E_{\rm median}$\\$({\rm TeV})$}
  & \entryc{$A$\\{$({\rm 10^{-5}\,erg\,cm^{-2}\,s^{-1}})$}}
  & \entryd{$\omega_1$\\$\omega_2$}
  & \entrye{$t_{\rm b,0}$\\$t_{\rm b,1}$\\$t_{\rm b,2}$\\$t_{\rm peak}$\\$({\rm s})$}
  & \entryf{$\alpha_0$\\$\alpha_1$\\$\alpha_2$\\$\alpha_3$}\\
  \midrule

  \entrya{$[30,\;+\infty)$}& \entryb{$0.54$}& \entryc{$2.22_{-0.16}^{+0.19}$}& \entryd{$1.07_{-0.15}^{+0.17}$\\$7.4_{-4.8}^{+4.3}$}& \entrye{$4.85_{-0.10}^{+0.15}$\\$15.37_{-0.88}^{+0.83}$\\$670_{-110}^{+230}$\\$18.0_{-1.2}^{+1.2}$}& \entryf{$14.9_{-3.9}^{+5.7}$\\$1.82_{-0.18}^{+0.21}$\\$-1.115_{-0.012}^{+0.012}$\\$-2.21_{-0.83}^{+0.29}$}\\
  \entrya{$[30,\;33)$}& \entryb{$0.35$}& \entryc{$1.81_{-0.20}^{+0.33}$}& \entryd{$1.46_{-0.46}^{+0.65}$\\$7.4$ (fixed)}& \entrye{$4.85$ (fixed)\\$16.1_{-1.8}^{+1.5}$\\$560_{-100}^{+160}$\\$17.9_{-2.2}^{+2.0}$}& \entryf{$12.8_{-5.4}^{+9.5}$\\$1.70_{-0.31}^{+0.46}$\\$-1.108_{-0.031}^{+0.029}$\\$-1.87_{-0.32}^{+0.23}$}\\
  \entrya{$[33,\;40)$}& \entryb{$0.41$}& \entryc{$2.63_{-0.42}^{+0.76}$}& \entryd{$0.85_{-0.26}^{+0.32}$\\$7.4$ (fixed)}& \entrye{$4.85$ (fixed)\\$13.4_{-2.2}^{+1.8}$\\$652_{-71}^{+68}$\\$17.0_{-3.2}^{+3.0}$}& \entryf{$5.9_{-2.4}^{+4.8}$\\$2.22_{-0.46}^{+0.76}$\\$-1.128_{-0.025}^{+0.023}$\\$-2.51_{-0.56}^{+0.40}$}\\
  \entrya{$[40,\;63)$}& \entryb{$0.54$}& \entryc{$2.43_{-0.27}^{+0.39}$}& \entryd{$1.01_{-0.22}^{+0.26}$\\$7.4$ (fixed)}& \entrye{$4.85$ (fixed)\\$15.1_{-1.5}^{+1.3}$\\$664_{-71}^{+79}$\\$18.0_{-2.0}^{+1.9}$}& \entryf{$20_{-6}^{+12}$\\$1.90_{-0.28}^{+0.39}$\\$-1.122_{-0.019}^{+0.018}$\\$-2.24_{-0.33}^{+0.23}$}\\
  \entrya{$[63,\;100)$}& \entryb{$0.81$}& \entryc{$1.84_{-0.16}^{+0.24}$}& \entryd{$1.70_{-0.45}^{+0.60}$\\$7.4$ (fixed)}& \entrye{$4.85$ (fixed)\\$17.8_{-1.5}^{+1.3}$\\$700_{-90}^{+110}$\\$18.8_{-1.8}^{+1.6}$}& \entryf{$7.8_{-2.5}^{+5.2}$\\$1.40_{-0.22}^{+0.29}$\\$-1.092_{-0.021}^{+0.020}$\\$-2.29_{-0.49}^{+0.29}$}\\
  \entrya{$[100,\;250)$}& \entryb{$1.62$}& \entryc{$1.79_{-0.35}^{+0.75}$}& \entryd{$0.98_{-0.40}^{+0.54}$\\$7.4$ (fixed)}& \entrye{$4.85$ (fixed)\\$15.4_{-3.7}^{+3.0}$\\$559_{-52}^{+55}$\\$18.6_{-4.9}^{+4.3}$}& \entryf{$13_{-5}^{+11}$\\$1.76_{-0.51}^{+0.52}$\\$-1.047_{-0.035}^{+0.031}$\\$-2.21_{-0.27}^{+0.23}$}\\
  \entrya{$[30,\;+\infty)$\\mask}& \entryb{$0.54$}& \entryc{$2.22_{-0.15}^{+0.19}$}& \entryd{$1.08_{-0.15}^{+0.17}$\\$7.4_{-4.8}^{+4.3}$}& \entrye{$4.85$ (fixed)\\$15.39_{-0.89}^{+0.81}$\\$670_{-110}^{+230}$\\$18.0_{-1.2}^{+1.1}$}& \entryf{$14.9_{-3.9}^{+5.7}$\\$1.82_{-0.18}^{+0.21}$\\$-1.115_{-0.012}^{+0.012}$\\$-2.21_{-0.83}^{+0.29}$}\\
  \entrya{$[30,\;+\infty)$\\high}& \entryb{$0.54$}& \entryc{$2.84_{-0.20}^{+0.24}$}& \entryd{$1.07_{-0.15}^{+0.17}$\\$7.6_{-4.9}^{+4.5}$}& \entrye{$4.85$ (fixed)\\$15.36_{-0.88}^{+0.83}$\\$660_{-90}^{+220}$\\$18.0_{-1.2}^{+1.2}$}& \entryf{$14.9_{-3.9}^{+5.7}$\\$1.84_{-0.18}^{+0.21}$\\$-1.108_{-0.012}^{+0.012}$\\$-2.19_{-0.81}^{+0.28}$}\\
  \entrya{$[30,\;+\infty)$\\low}& \entryb{$0.54$}& \entryc{$1.75_{-0.12}^{+0.15}$}& \entryd{$1.08_{-0.15}^{+0.17}$\\$7.2_{-4.7}^{+4.1}$}& \entrye{$4.85$ (fixed)\\$15.38_{-0.87}^{+0.82}$\\$680_{-90}^{+230}$\\$17.9_{-1.2}^{+1.1}$}& \entryf{$14.9_{-3.9}^{+5.7}$\\$1.81_{-0.18}^{+0.21}$\\$-1.123_{-0.012}^{+0.012}$\\$-2.23_{-0.86}^{+0.29}$}\\

\bottomrule
\end{tabular}
\end{center}
\end{table}

\end{document}